 \documentclass[final,3p,authoryear]{elsarticle}

\usepackage{ulem}
 \usepackage{color}
 \usepackage{natbib}

\newcommand{\hMpc}{{\ifmmode{h^{-1}{\rm Mpc}}\else{$h^{-1}$Mpc }\fi}}  
\newcommand{\hGpc}{{\ifmmode{h^{-1}{\rm Gpc}}\else{$h^{-1}$Gpc }\fi}}  
\newcommand{\hmpc}{{\ifmmode{h^{-1}{\rm Mpc}}\else{$h^{-1}$Mpc }\fi}}  
\newcommand{\hkpc}{{\ifmmode{h^{-1}{\rm kpc}}\else{$h^{-1}$kpc }\fi}}  
\newcommand{\hMsun}{{\ifmmode{h^{-1}{\rm {M_{\odot}}}}\else{$h^{-1}{\rm{M_{\odot}}}$}\fi}}  
\newcommand{\hmsun}{{\ifmmode{h^{-1}{\rm {M_{\odot}}}}\else{$h^{-1}{\rm{M_{\odot}}}$}\fi}}  
\newcommand{\Msun}{{\ifmmode{{\rm {M_{\odot}}}}\else{${\rm{M_{\odot}}}$}\fi}}  
\newcommand{\msun}{{\ifmmode{{\rm {M_{\odot}}}}\else{${\rm{M_{\odot}}}$}\fi}}  

\newcommand{\kms}{\ {\rm km\ s^{-1}}}

\newcommand{\lcdm}{$\Lambda$CDM}

\journal{New Astronomy Reviews}

\begin{document}

\begin{frontmatter}



\title{Dark Matter in the Local Universe}


\author{Gustavo Yepes}

\address{Departamento de F\'\i sica Te\'orica M-8, Universidad Aut\'onoma de
  Madrid,  Cantoblanco 28049 Madrid Spain}

\author{Stefan Gottl\"ober}

\address{Leibniz Institut f\"ur Astrophysik, An der Sternwarte 16, 14482
  Potsdam, Germany}

\author{Yehuda Hoffman}

\address{Racah Institute of Physics, The Hebrew University of Jerusalem, 91904 Givat Ram, Israel}

\begin{abstract}
We  review  how dark matter is distributed  in our local
neighbourhood from an  observational and theoretical perspective. We
will start by describing first the  dark matter  halo of  our own
galaxy and in the Local Group.  Then we  proceed to describe the  dark matter
distribution in the more
 extended area known as the Local Universe.  
Depending on the nature of dark matter, numerical simulations 
 predict  different abundances of  substructures in Local Group
 galaxies, in the number of  void regions  and  the abundance of  low rotational velocity  galaxies in the Local Universe.
 By comparing  these predictions with
the most recent observations,  strong constrains on the
physical properties of the dark matter particles can be
derived.  We devote particular attention to the results from  the  
Constrained Local UniversE Simulations (CLUES) project, a special set of simulations 
 whose initial  conditions are constrained by observational data from the Local Universe. 
The resulting simulations are designed to reproduce 
the observed structures  in the nearby universe. The CLUES provides a numerical laboratory
for simulating the Local Group  of galaxies and exploring  the physics
of galaxy formation in an environment designed to follow the observed Local
Universe. It has come of age as the numerical analogue of Near-Field Cosmology.
\end {abstract}


\end{frontmatter}



\section{Introduction}
\label{intro}


It is  widely attributed to Fritz \citet{Zwicky}    the   introduction of the name  {\it dark matter} (DM)  to account for the  non-visible
 mass   required  to explain the large velocity dispersion  of  the galaxies  in the Coma cluster that he derived by applying the Virial Theorem .  But, in fact, he was  referring   to the term used first by the Dutch astronomer Jaan \citet{oort} one year earlier.   Oort  studied the dynamics of the brightest stars  in the disk of the Milky Way (MW). From his   analysis   he deduced that the  total density exceeds
 the density of visible stellar populations by a factor of up to 2. This limit is often called the Oort limit. Thus, he concluded that the amount of invisible matter in the Solar vicinity could be approximately equal to the amount of visible matter. He named it,  for the  first time,  this invisible component as { \it dark matter}.    Later on,   \citet{babcock}  measured, for the first time, the mass distribution  in the Andromeda Galaxy (M31) from the radial velocity curves derived from optical emission line regions. He concluded that the mass  profile of M31 monotonically increases from the centre  outwards, up to 20 kpc,  the maximum distance he could observe.  In the late 50\rq{}s, the  first 21-cm observations  \citep{hulst},  corroborated the  earlier optical results from Babcock and clearly indicated that  the rotation curve of M31 flattens off  at around 35 kpc with no indication of a decay.  During the  late 70\rq{}s 
\citep{bosma}
and early 80\rq{}s, \citep{rubin},   more  observations of  M31 and other spiral galaxies clearly indicated that their rotation curves were  flat out to large distances from the optical emission 
\citep[see e.g.][]{einasto_review} for a more detailed description of the historical development of the DM concept). 

A completely different, more theoretically based,  approach in the determination of  the  total masses of   MW and M31 was used by  \citet{kahn-woltej}.  The evidence that  M31 has a negative redshift  of about $120 \kms $  towards our Galaxy  can be explained, if both galaxies, M31 and MW, form a gravitationally bound system. A negative radial velocity indicates that these galaxies have already passed the apocenter of their relative orbit and are
presently approaching each other. From the approaching velocity, the mutual distance, and the time since passing the pericenter (taken as the present age of the Universe) the
authors calculated the total mass,   assuming a two body point-like system.  They
found that the combined mass of M31 and MW is $M_{tot} \ge 1.8\times 10^{12} \Msun$.  This value is a factor of $\sim $ 10 higher than the conventional mass estimates of the two galaxies ($\sim 2\times10^{11} \Msun$).  This method is known as the {\it Timing Argument}  and it is one of the first observational evidence  that the  total gravitating mass   of the Local Group (LG) exceeded the visual  one by almost an order of magnitude.   Nevertheless,  the hypotheses  in which the Timing Argument is  based  have not been  tested  until  recently, when simulations can make possible to  trace  back the formation history of  LG-like objects. We will come back to this in  \S  \ref{subset:TA}.

There is now  an overwhelming   amount of observational evidence, at many different scales,  that  firmly supports the  idea that  there exists much more matter in the Universe than
just the luminous matter.  The ratio between the dark and the visible
matter components  grows with  scale.  It is widely assumed  that gravitational instability of the primordial density perturbations  in the collisionless  DM fluid is the main mechanism that drives the formation of structure.  
The  standard $\Lambda$ Cold Dark Matter ({\lcdm})-model, where $\Lambda$ is the Cosmological Constant, 
of  cosmological structure formation describes
very well the observations of  the large scale structure (LSS),
\citep[see e.g.][for a review]{frenkwhite2012}.

At present,  the parameters of
the model have been determined  with very high precision
\citep{2013arXiv1303.5076P}.  Despite all the success of the model   to explain LSS, the
formation of small scale structure seems to be an open problem.  It has
been  known for a long time that the model predicts more small scale structures
than observed \citep
{1999ApJ...522...82K,1999ApJ...524L..19M,2007ApJ...657..262D,2008MNRAS.391.1685S}. 

Numerical N-body simulations  have  given   us   a clear picture of how DM   is structured at different scales.  At large scales, DM is distributed   in the universe  in the form of a web, the so called {\it cosmic web} \citep{1996Natur.380..603B}. 
Observationally,  the  distribution of  galaxies  in the universe,
 as well as the distribution of total matter, as inferred from its gravitational lensing and
reconstructions from large galaxy surveys give  also the appearance that 
mass and light are distributed in a web-like structure dominated by
linear filaments and concentrated compact knots, thereby leaving
behind vast extended regions of no or a few galaxies and of low
density  \citep{2010MNRAS.409..355J,2011MNRAS.417.1303M,2012MNRAS.420.1809W,
2012MNRAS.425.2422K}. 
Direct mapping of the mass distribution by weak lensing reveals
a time evolving loose network of filaments, which connects rich
clusters of galaxies \citep{2007Natur.445..286M}. The extreme
low resolution of the weak lensing maps cannot reveal the full
intricacy of the cosmic web, and in particular the difference
between filaments and sheets, yet they reveal a web structure
that serves as a gravitational scaffold into which gas can
accumulate, and stars can be built.

Our  Local Universe is the  best place to  test the predictions of the
{\lcdm}  model  down to the smallest scales given by the free
streaming  of the DM particles.  Therefore,   the Local Universe  can
be considered to be a cosmic laboratory to attempt to identify the
nature of DM.   In  the past years there has been an enormous
experimental effort devoted to  identify  the particle physics
candidate for DM  \citep[see e.g.][for a recent review]{strigari}. 
Underground direct detection experiments try
to find  the signature of the elusive dark matter particles  of the
galactic halo when they weakly interact with the nuclei of detector's
material (noble gas, crystalline salt, semiconductors, etc).  On the
other hand, the FERMI satellite is now searching for the gamma photons
coming from  disintegration or annihilation of the dark matter
particles,    at the
Milky Way centre,   in   its dwarf galaxy satellites, or even in
extragalactic sources like  the nearby Andromeda Galaxy or   in Virgo, 
the closest  galaxy cluster to us.  The basic
hypothesis behind all these experimental efforts is that the
constituent of dark matter is  a non-baryonic  Weakly Interacting  Massive
Particle (WIMP) like the neutralino,   that is predicted  in   Supersymmetry theories.

On the other hand, there are other probes that try to measure the level
of structures formed by gravitational growth of  DM
fluctuations at  small scales, where the predictions for abundance of
low mass objects strongly depend on the individual mass of the  DM particles.
  In this regard,  the number of satellite galaxies
around our Milky Way, or the number of low mass HI galaxies in our
local neighbourhood are two excellent  observational tests for dark
matter models.  But, a direct comparison between observations and
theoretical predictions must be done with caution. The dynamics of our
Local Universe has some special features  due to the peculiar
distribution of the matter around us. Any realistic simulation  of
structure formation in a particular dark matter model 
 should account for these features before a  reliable comparison with
 observations  can be made. Otherwise, the comparison could be biased
 due to the cosmic variance.
 
       There have been some attempts  to
 minimize the effect of cosmic variance  by simulating the formation of
 cosmological  structures which are designed to resemble our own Local   Universe.
 The so-called CLUES (Constrained Local UniversE Simulation, 
 {\tt http://clues-project.org}  )  collaboration
 is trying to do so by  imposing observational constrains on the
 otherwise random realizations of a cosmological initial  
density perturbation field. As a result, the structures formed 
reproduces the main features of the observed   most  massive clusters and
superclusters such as Coma, Virgo or  the Local Supercluster.  
Thus, LG-like objects are formed in an environment that resembles the real one. In this  context, the CLUES Local Groups  can
be considered as  numerical  proxies that can serve to study issues such as:
\begin{itemize}
\item how typical our LG  is  and what
can be  learnt from it about structure formation at large; 
\item the structure of the stellar halos in the LG; 
\item tidal streams and 
the formation history of the LG; 
\item the missing satellites problem; 
\item  how   does the baryon physics 
affect the Dark Matter  distribution; 
\item  the nature of nearby dwarf galaxy associations beyond the LG;
\item  improved predictions for Dark Matter  
detection; 
\item galaxy formation and environmental dependence in the
framework of the cosmic web; 
\end{itemize}

In the upcoming era of the GAIA\footnote{\texttt{http://www.gaia.esa.int}} and
PANDAS\footnote{\texttt{https://www.astrosci.ca/users/alan/PANDAS/‎}}
observations of the LG, 
simulations like those performed within the framework of  CLUES are the  numerical counterparts.
The combined high quality observations and detailed simulations will shed
new light on the formation history of the LG and will provide a new
framework for understanding its  cosmological implications.    

In this review we  try to summarize the main results achieved within   the CLUES collaboration during the past years    on several of the above mentioned items. The ``L"  in the CLUES stands for ``Local"  yet local is not a well defined concept. Throughout the paper we use the term  ``Local"  to denote a finite region of the universe which harbours the LG close to its center and  that extends over linear scales ranging typically  from a very few to a few tens of Megaparsecs.

The paper is structured as follows: 
In \S \ref{clues}  we give a brief description of the CLUES project and summarize the numerical experiments that have been done so far.  Then  in \S \ref{mw}  we  focus on  the  study of the dark matter distribution  in the Local Group and how baryon physics affects the distribution of dark matter, which is of extreme importance for  experiments of dark matter detection.   In \S \ref{subset:LG} we  review the main results from CLUES simulations on the formation histories of the Local Group and how unique the LG  is as compared with  other binary systems  formed in dark matter N-body simulations.  We continue  in \S \ref{weight} with a discussion of the  estimation of the total mass of LG based on different mass estimators, including the timing argument, and how well these  work on simulations.  We  move  to larger scales and describe  in \S \ref{cosmicweb} how the cosmic web of dark matter   in the Local Universe   can be used  for Near-Field Cosmological studies. In  \S \ref{lu} we show    how the Local Universe can also be used as a cosmic laboratory to discern among  different candidates to dark matter. We  conclude in \S \ref{summary} with a summary of the  main results  presented in this review.

\section{Simulating the Local Universe: The CLUES project}
\label{clues}

The Local Universe is the best observed part of the universe in which least  massive and  faintest  objects can be detected and studied in detail.  These
observations resulted in  a new research field  called {\it Near-Field Cosmology } 
and have motivated cosmologists to study the  LG  archaeology 
in their quest for understanding galaxy formation and  the play dark matter has on it.
This also motivated the CLUES collaboration 
to perform a series of numerical simulations of the evolution of the local
universe. For these simulations we constructed the initial conditions based on
the observed positions and peculiar velocities of galaxies in the Local
Universe. These simulations reproduce the local cosmic web and its key
players, such as the Local Supercluster, the Virgo cluster, the Coma cluster,
the Great Attractor and the Perseus-Pisces supercluster.  Such constrained
simulations cannot directly constrain small scale structure on sub-megaparsec
scales, yet they enable the simulation of objects  on these scales within the
correct environment. Therefore,  such simulations provide a very attractive possibility
of simulating the Local Group of galaxies within the right environment.  We
have used these constrained simulations as a numerical Near-Field Cosmological
Laboratory for experimenting with the complex gravitational and gasdynamical
processes that leads to the formation and evolution of galaxies like our own MW and its neighbour,  M31.

\subsection{ Observational Data}
\label{observation}
 
Observational data of the nearby universe is used as constraints on the
initial conditions and thereby the resulting simulations reproduce the
observed large scale structure.  The implementation of the \cite{1991ApJ...380L...5H}  algorithm of
constraining Gaussian random fields  to  follow observational
data and a description of the construction of constrained simulations can be found in  detail in \cite{2002ApJ...571..563K} and  in \cite{2003ApJ...596...19K}. 
Here,  we briefly describe the observational data used so far.  

To set up the  CLUES initial conditions
we used radial velocities of galaxies drawn from the MARK III
\citep{1997ApJS..109..333W}, SBF \citep{2001ApJ...546..681T} and the
\citet{2004AJ....127.2031K} catalogues. Peculiar velocities are less
affected by non-linear effects and are used as constraints as if they were
linear quantities \citep{1999ApJ...520..413Z}.  The other constraints are obtained
from the catalogue of nearby X-ray selected clusters of galaxies
\citep{2002ApJ...567..716R}.  The data constrain the simulations on scales larger
than $\approx 5 \hMpc$ \citep{2003ApJ...596...19K}. It follows that the main features
that characterize the Local Universe (e.g.  Virgo, Local Supercluster, Coma, Great Attractor, etc)
are all reproduced by the
simulations. The small scale structure is hardly affected by the constraints
and is essentially random.

Currently,  the Cosmic Flow 2 survey with more than 8300
galaxies which extends to 12,000 km/s with a  median error on distances of $\sim15$\%
\citep{2012AN....333..436C,2012ApJ...749..174C,2012ApJ...749...78T,2013AJ....146...86T} is used to set up initial conditions for the next generation CLUES simulations.

\begin{figure}[ht]
\begin{center}
\includegraphics[width=0.8\textwidth]{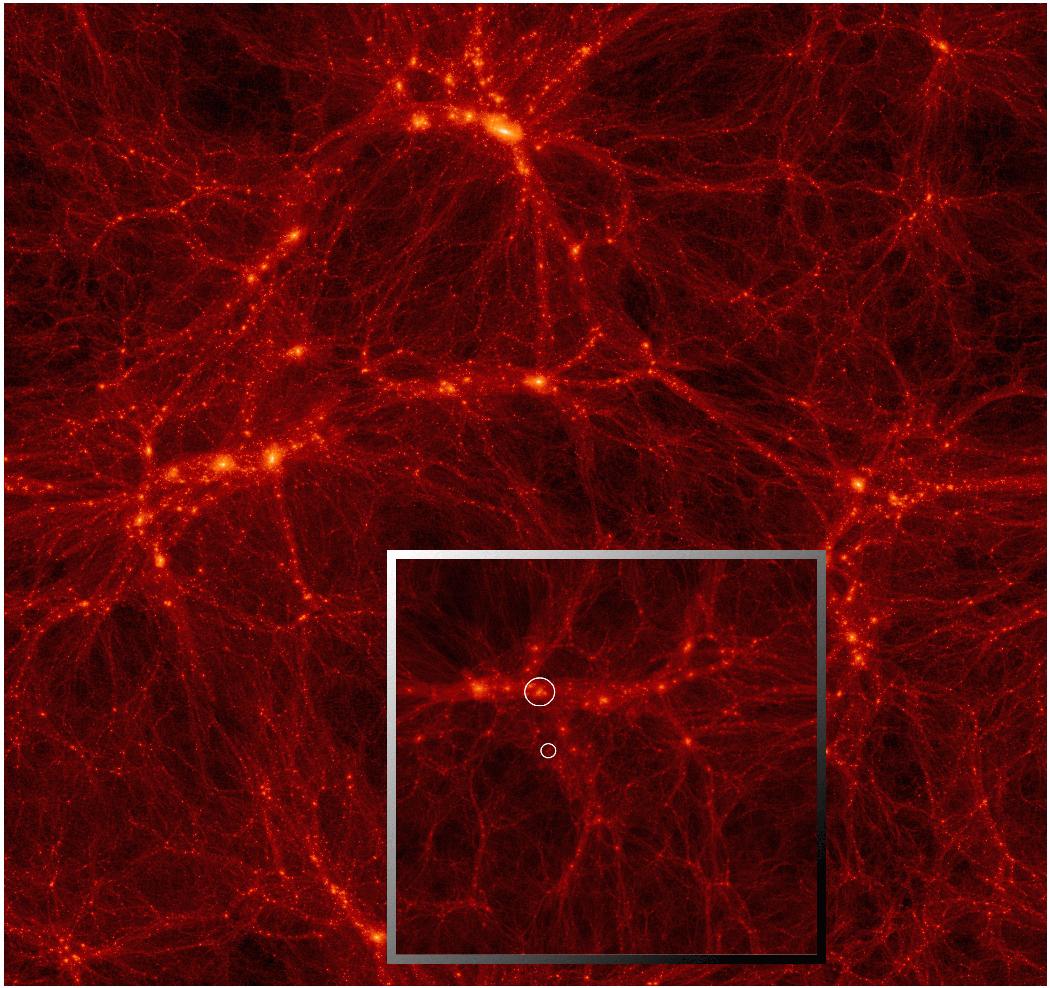}
\caption{ Dark matter  density in two constrained simulations  in  a box of 160 \hMpc size and 64 \hMpc
  size (smaller inset) with different underlying random realizations}  
\label{fig:64to160}
\end{center}
\end{figure} 

\subsection{Constrained Initial conditions}
\label{subsec:IC}

The Hoffman-Ribak algorithm is used to generate
the initial conditions as constrained realizations of Gaussian random
fields on a $256^3$ uniform mesh, from the observational data
mentioned above.  Since these data only constrain scales larger than a
very few Mpc, we have  performed  a series of different realizations in
order to obtain one which contains a LG  candidate with the correct
properties (e.g. two halos with proper position relative to
each-other, mass, negative radial velocity, etc).  
High resolution extension of the low resolution constrained
realizations were then obtained by creating an unconstrained
realization at the desired resolution, FFT-transforming it to $k-$space
and substituting the unconstrained low $k $ modes with the constrained
ones. The resulting realization is made of unconstrained high $k$ modes
and constrained low $ k$ ones. 

The constrained simulations performed so far do not account for the shift of
the objects with respect to the 
unperturbed background. 
Using the  Reverse Zeldovich Approximation for
constructing the initial conditions 
improves 
the quality substantially \citep{2013MNRAS.430..902D,2013MNRAS.430..912D,2013MNRAS.430..888D} 
and  it is  being used in the next generation CLUES simulations.

\subsection{Constrained  Simulations of the full box}
\label{subsec:simulation}

Using the above initial conditions, we carried out the simulations using the publicly available N-body + SPH code  \textsc{Gadget2}  \citep{2005MNRAS.364.1105S}.   

Two different computational volumes have been  used. To study the structures in the  Local Universe, a box of  $160$  \hMpc was simulated. This box is nevertheless, too big  to be able to study in detail the  internal structure of LG-like objects. Therefore,  for the study of the LG and the Local Volume (few Mpc around the LG)
a smaller computational box of 64  \hMpc was used.

In Fig. \ref {fig:64to160} we compare  the DM distribution of the constrained
simulations in the  two  computational volumes. 
 As mentioned above,  the small scale structure is added in these
simulations by random modes according to the underlying cosmological
model.  The large plot corresponds to the DM distribution in the 160 \hMpc volume and  the inset  plot  shows the DM  in   the  smaller box of 64 \hMpc.  The two simulations use the same observational
constraints but completely different random phases for the remaining small
scale perturbations. Nevertheless it is impressive how well the large scale
structure - the Local Super-cluster - is reproduced in both simulations (see Table \ref{tab:fullbox} for further details).

To find the  LG we first identify  the Virgo cluster (the large circle in Fig.  \ref {fig:64to160})  . Then we search for an object which
closely resembles the LG and is in the right direction and
distance to Virgo  (the small circle in Fig.  \ref {fig:64to160}).  Since 
the small scale structure is unconstrained, 
the only  possibility to obtain a LG-like object is to produce many different realizations  with the same constrains. Such procedure has yielded 3 good LG-like objects out of 200 constrained initial conditions. 

\begin{figure}[ht]
\begin{center}
\includegraphics[width=0.8\textwidth]{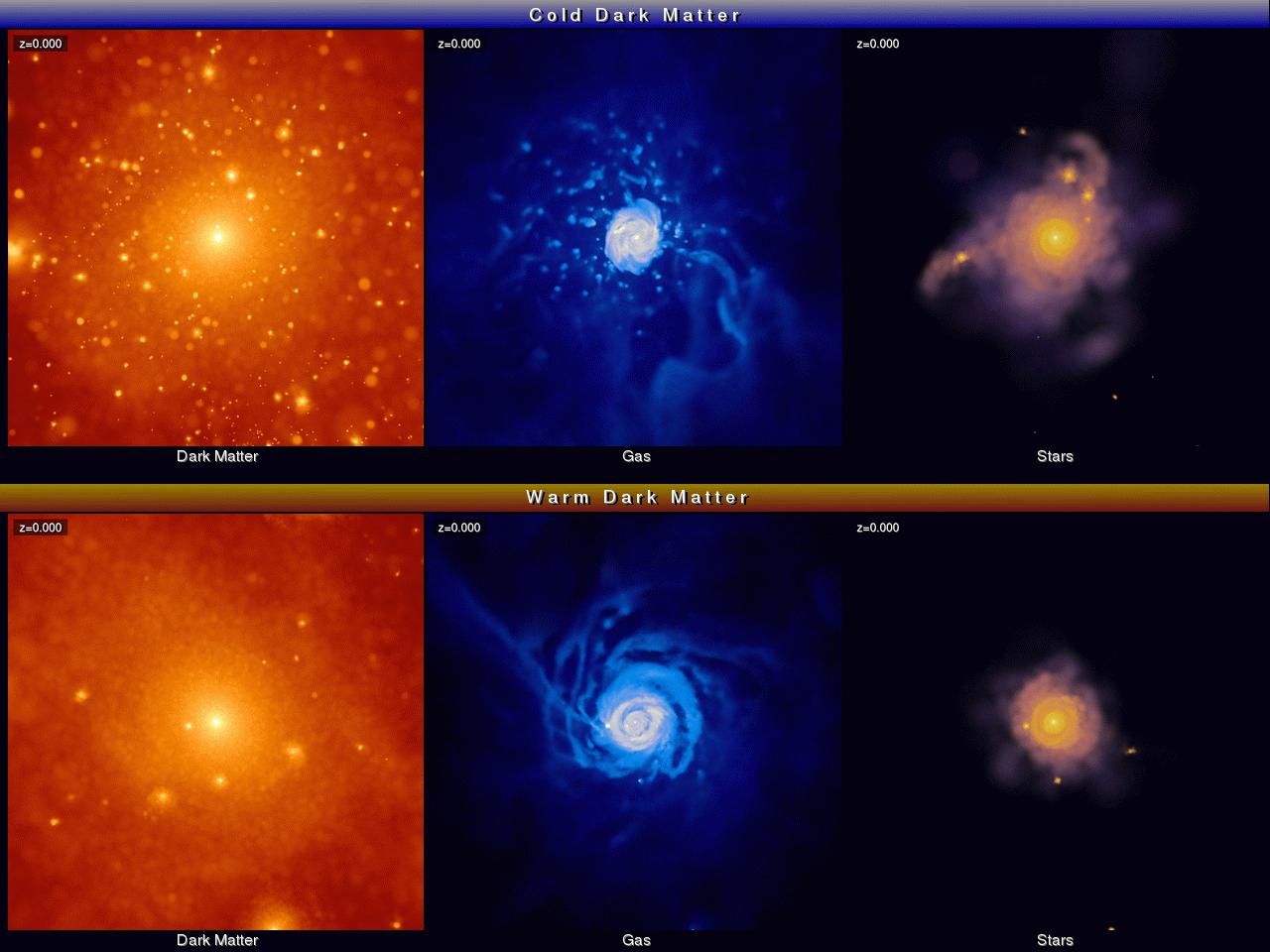}
\caption{ Comparison of a CDM and WDM simulations of a galaxy in
  the Local Group (Top: CDM, Bottom: WDM; from left to right DM, gas,stars) }
\label{fig:CDM_WDM_LG}
\end{center}
\end{figure}

\subsection{ Zoomed Simulations of the Local Group}
\label{subsec:CRzoomed}

In order to study the evolution of the LG in more detail we performed
zoomed simulations of the evolution of the Local Volume. To this end we have
identified spherical regions around the LG candidate  at redshift $z=0$ and used
initial conditions with higher resolution in this region. They were constructed following the prescription set out in
\citet{2001ApJ...554..903K}.   The main idea is to keep high resolution in the
sphere of interest and to decrease the resolution in shells with increasing
radius up to a low resolution ($256^3$ particles) in the rest of the box. By
construction we keep the same phases so that the high and low resolution
simulations can be directly compared. 

In some of the CLUES simulation we replace,   in the high resolution area,  the Dark
Matter particles by pairs of DM and gas particles and follow their evolution
using the  entropy-conserving  SPH version of the \textsc{Gadget2} code \citep{2002MNRAS.333..649S} in order to
reduce numerical overcooling.  Assuming an optically thin primordial mix of
hydrogen and helium the radiative cooling is computed following
\citet{1996ApJS..105...19K} and photoionization by an external uniform UV
background is computed following \citet{1996ApJ...461...20H}. Finally,  star formation is produced from a   two-phase interstellar medium of hot and
cold gas clouds  using a subgrid model \citep{1997MNRAS.284..235Y,2003MNRAS.339..289S}.   Including
gas-dynamical processes in the simulation one can show that also the DM distribution  is changed as well.

As an example of  what  the CLUES  gasdynamical  simulations look like,  we compare in the upper and lower part of
Fig. \ref{fig:CDM_WDM_LG} a 
Cold and a Warm Dark Matter (CDM and WDM,  respectively) simulation of a galaxy
in the LG object.   On the left panel of this figure,  the DM  distribution is
shown. The middle panel shows the gas and  the stars are in the right one. 
 The gaseous  and the stellar disks can be clearly seen.

In Tables \ref{tab:fullbox} and \ref{tab:resimu} of  the Appendix  we  summarize the main features of the  CLUES simulations done so far.

\section{Dark Matter  in the  Local Group}
\label{mw}

\begin{figure*}
\includegraphics[width=8cm]{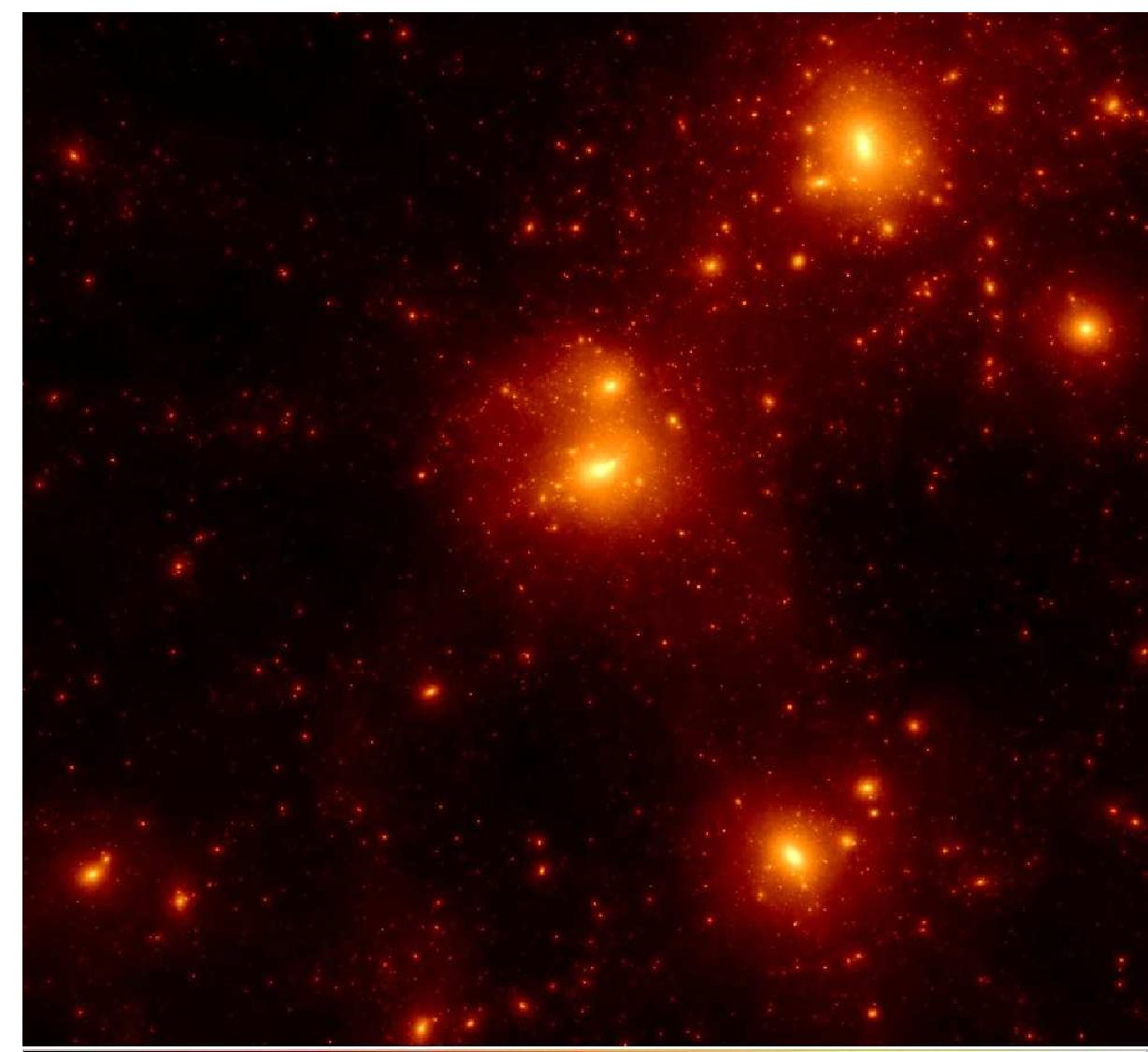}
\includegraphics[width=8cm]{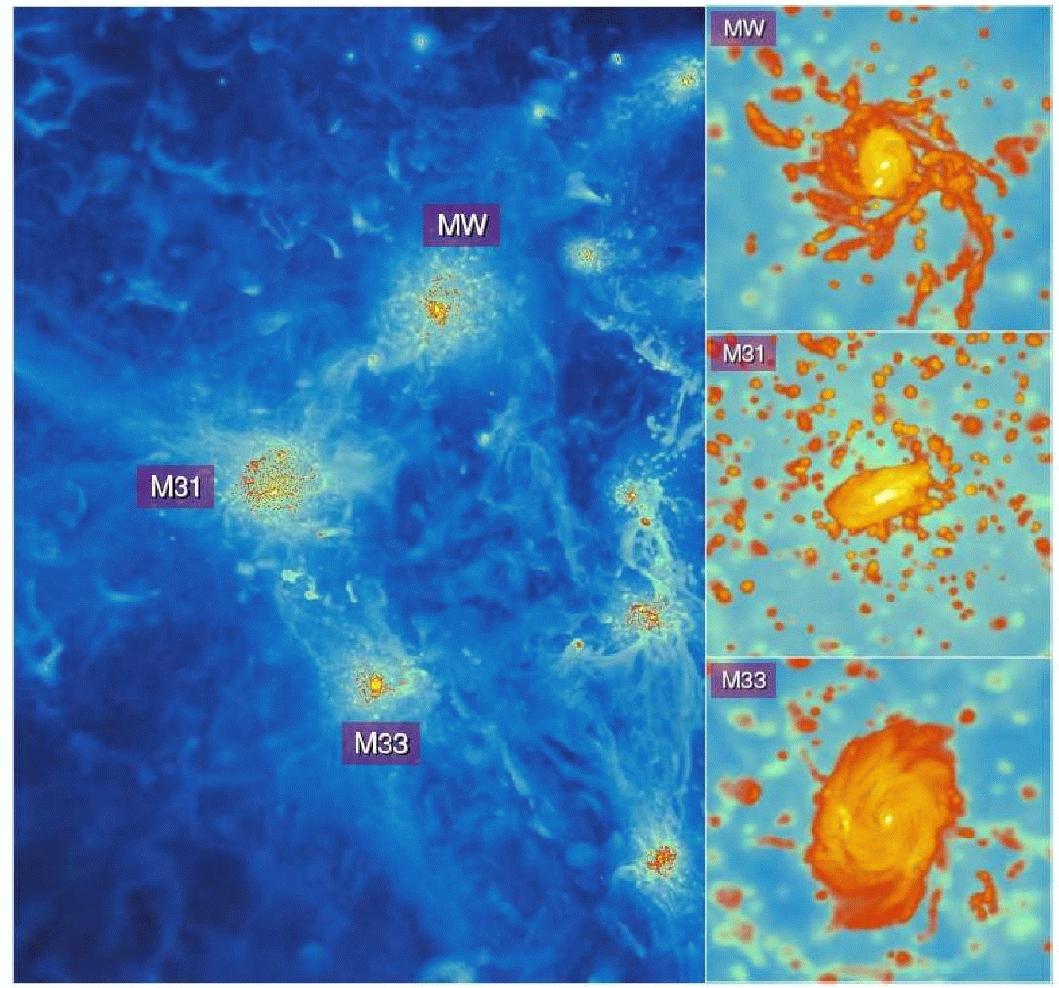}
\caption{ The dark matter distribution of the Local Group candidate found in the dark matter only 
 LG64-3 CLUES simulation (Left) and the gas distribution    in the same object   from the LG64-3 gasdynamical SPH  simulation.  The area is approximately 2 \hMpc  across.  }
\label{fig:wmap3_lg}
\end{figure*}

The  observational evidence that the rotation curves of M31 and other spirals  are flat  suggest that the radial distribution of total matter (stars, gas and dark matter)  follow a near isothermal profile with $\rho(r) \propto r^{-2} $. Since dark matter is the dominant component, it should also follow such kind of profile.  In the early  90s
the N-body simulations showed that CDM halos  do follow  a rather universal density profile parametrized by  the so-called  NFW formula \citep{nfw}.

\begin{equation}
\rho_{NFW}(r) = \frac{4\rho_s}{\frac{r}{r_s}(1+\frac{r}{r_s})^2}
\label{eq:nfw}
\end{equation}
where $\rho_s$   and $r_s$ are  characteristic  density and radius.  
This fit  presents a singularity at $r\rightarrow 0$, although the total integrated mass is finite.  This sharp rise of the density at the halo centre   forms a \lq\lq{}cusp\rq\rq{}.  The NFW has been generalized   to allow for different values of the asymptotic slopes  towards the centre and to  the outskirts.

\begin{equation}
\rho(r) = \frac{ 2^{(c-\alpha)/\beta} \  \rho_s}{\left(\frac{r}{r_s}\right)^\alpha \left (1+\left(\frac{r}{r_s}\right)^\beta \right )^{(c-\alpha)/\beta}}
\label{eq:nfwgen}
\end{equation}
giving the possibility to  fit  profiles that are  cuspier  ($\alpha >1$ )  or  cored  ($\alpha <1 $) .  The core-cusp problem has been a subject of many recent studies, based both on observational data as well as on results from  very high-resolution N-body  simulations (see eg. \citet{deblok} for a  review).   The latest numerical results have favored    another 
 kind of fitting formula  for the density profile of  dark matter
 halos  \citep{navarro2010,2013MNRAS.431.1220D}
\begin{equation}
\rho_{\rm E}(r)= \frac{\rho_{-2}}{e^{2n\left[\left(\frac{r}{r_{-2}}\right)^\frac{1}{n}-1\right]}}
\label{eq:einasto}
\end{equation}
where $r_{-2}$ is the radius where the logarithmic slope of the density profile equals -2 and $n$  is a parameter that describes the shape of the density profile. 
The $r_{-2}$  scale radius is equivalent to    $r_s$ of the NFW profile, %
and the density $\rho_{-2}=\rho(r_{-2})$ is related to the NFW one through $\rho_{-2}=\rho_s/4$.  This profile gives also a  finite total mass, but  its logarithmic slope decreases inwards more gradually than a NFW, with no  asymptotic slope at the centre. This profile is known as the  Einasto profile, since it was first proposed by Jaan Einasto in 1965  to model the kinematic of stellar systems.

Knowledge on the DM distribution in the Local Group, and beyond, is essential to any DM detection experiment. This is certainly valid for the indirect detection channel of gamma rays from annihilation/disintegration  of DM particles. Knowledge of the full phase space distribution of the DM particles in the Solar neighbourhood is critical for the proper interpretation of direct detection terrestrial experiments. However, this knowledge of the DM distribution is astronomically severely hindered. Despite the fact that the MW galaxy is the best studied of all galaxies, our position inside  one of its  spiral arms makes 
it difficult to  study accurately the total mass distribution,  compared with external galaxies. This situation can be remedied by the use of numerical simulations and experiments.

There are different methods to derive the  total gravitating mass of our Galaxy, depending on the distance from us  (see eg. \citet{strigari}).   Within the central few parsecs, the contribution of dark matter and the disk stars is not very important. The mass is dominated by the bulge and the central black hole, which is estimated to  have a mass of $4\times 10^6 $ \Msun.  Assuming that the dark matter follows a  NFW profile, the integrated  mass  of dark matter is no more than a few 1000 \Msun within the few kpc around the centre.  Thus, it is not possible,  with the current observations,  to  derive the asymptotic slope of the dark matter  mass profiles. It is expected, from results  of numerical simulations, that the density  profile may be steeper and cuspier  than the one given by NFW fit due to  the 
adiabatic compression of the DM in response to the collapse of baryons to the centre \citep{gnedin2011}.  

\begin{figure}[ht]
\includegraphics[width=16cm]{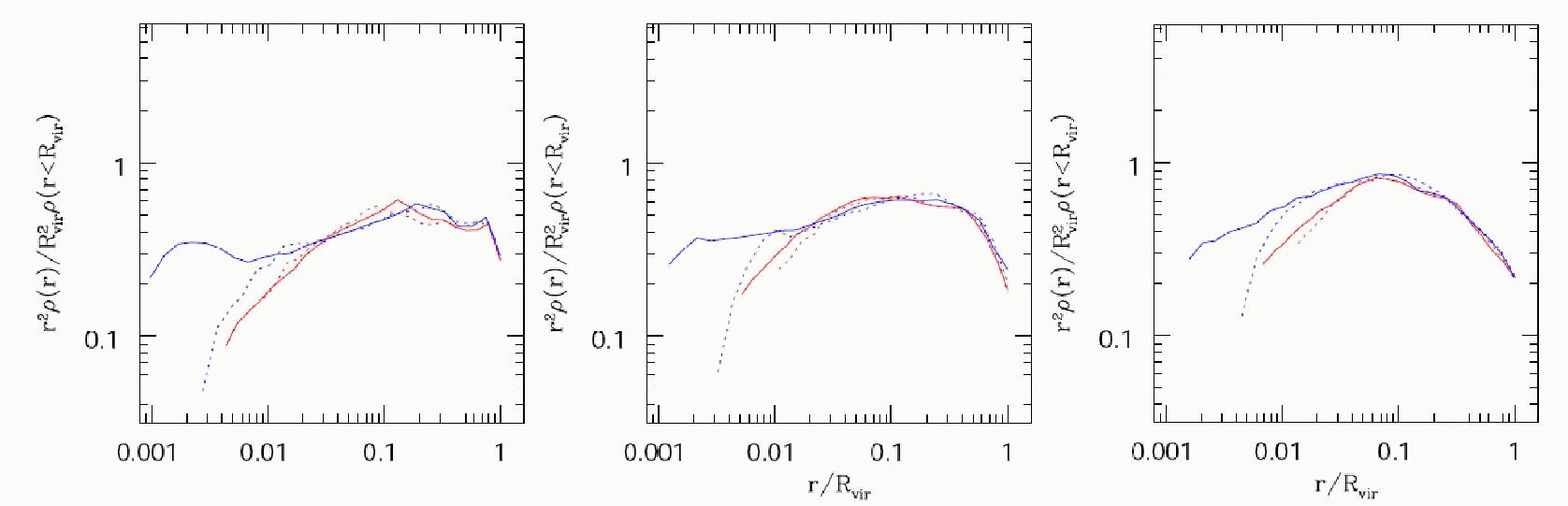}
\caption{The DM density profiles, scaled by $r^2$,  for the three main halos of the LG found in the LG64-3  CLUES simulation (see Table \ref{tab:resimu}). 
From left to right: M31, MW and M33, named according to the descending masses.  The results from the dark matter only simulation are shown in solid  red while the same  objects simulated with baryons  (gas and stars) are shown in solid  blue. The dotted lines represent the results from  simulations with 8 times less resolution in mass.  (Figure taken from  Luis A. Martinez-Vaquero's PhD thesis)  }
\label{fig:profiles}
\end{figure}

Fig. \ref{fig:profiles}  shows the DM  density profile   of  the three most massive galaxies (MW, M31  and M33, hereafter)   of the  
LG64-3-DM and LG64-3-SPH  high-resolution  CLUES simulations (see Table \ref{tab:resimu}).  
As can be seen,   there are substantial  differences in the slopes  of
the dark matter when  baryons  and their physical effects (cooling, star formation, etc) are taken into account.  While in   the collisionless DM only simulations the profiles are well in agreement with NFW,  the DM profiles in the gasdynamical run   have  steeper slopes, closer to -2 at the inner parts.  This effect on the DM  caused by the cooled baryons  that sink to the centre of the DM halos is known as  ``adiabatic contraction" and it can be modelled analytically \citep[and references therein]{gnedin2011},  giving a reasonable good approximation to the   simulation results.  The inclusion of this effect is crucial for a correct estimation of the DM density at the galactic centre and can make substantial differences  in the estimation of the gamma ray fluxes   coming from  dark matter annihilation \citep{vargas2013}.

Another effect that can be observed when physical processes involving baryons are considered in the simulations is that the shape of the DM halo is different from
that obtained in simulations containing only DM. We have also quantitatively studied
this effect by calculating the eigenvalues and eigenvectors of the tensor of inertia at different
distances from the Galactic Centre in  the  
 CLUES galaxies. 
\begin{figure}[ht]
\includegraphics[width=16cm]{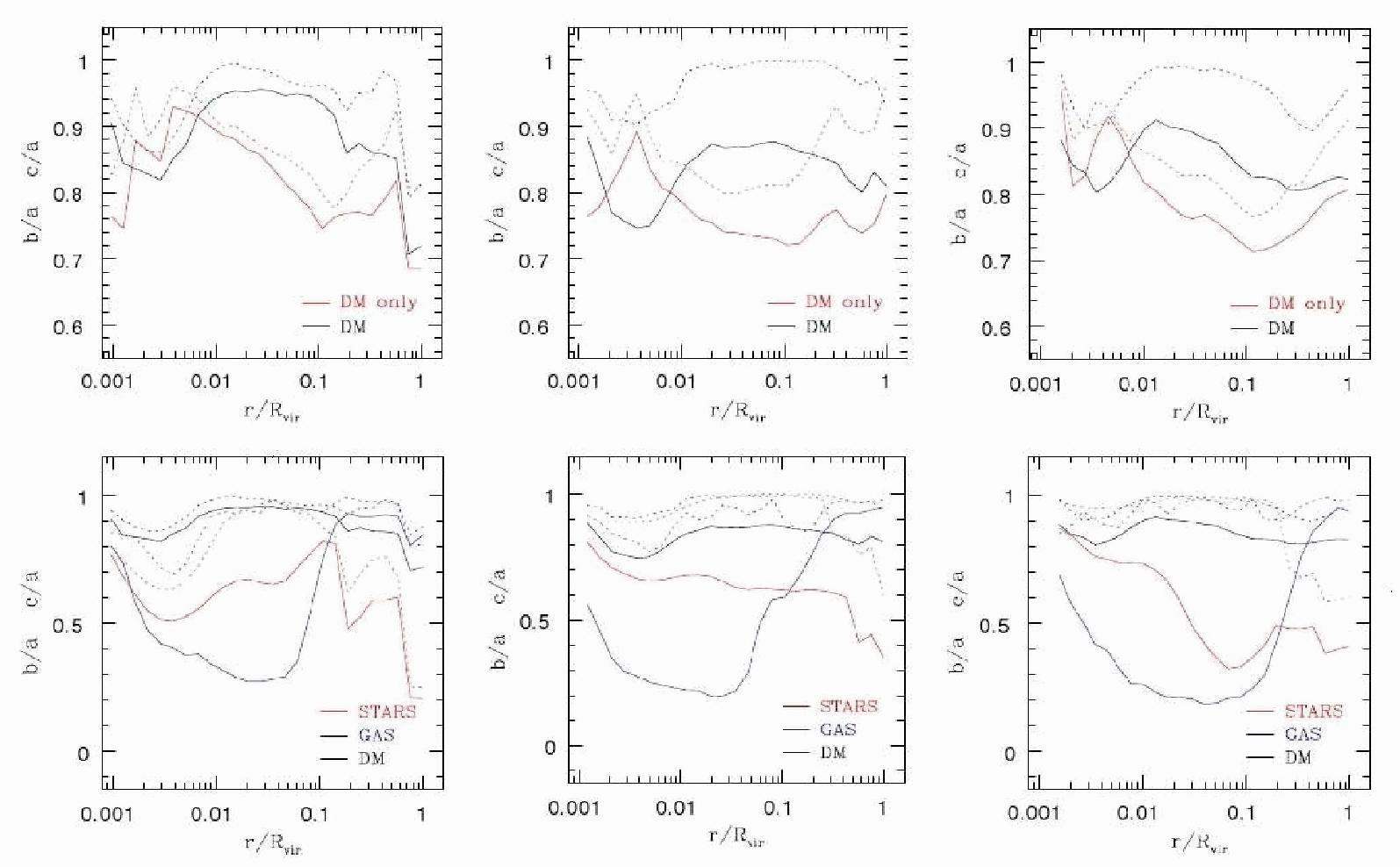}
\caption{Ratios of the  3 eigenvalues  ($a>b>c$) of the inertia tensor (solid curves represent $c/a$, and dotted  represent  $b/a$),  corresponding to the distribution of stars, gas and DM  for the three main halos of the LG found in the LG64-3 CLUES simulation.  From left to right: M31, MW and M33, named according to the descending masses.  The results from the DM-only simulation is shown in the upper row, while the results for the gasdynamical simulation is shown in the lower row. (Figure taken from  Luis A. Martinez-Vaquero's PhD thesis) }
\label{fig:axis}
\end{figure}
 Fig. \ref{fig:axis}  shows 
  that  the  ratio of the eigenvalues of the inertia tensor ($a >b
  >c$)   approaches  1, indicating that  the  DM distribution in MW
  type halos  becomes more spherical  when baryons are taken into
  account. These results gives theoretical support to the common
  assumption   of spherical symmetry  used in models that try to
  reconstruct the  mass distribution of the MW from a variety of
  observations \citep[e.g.][]{ullio}. From observations,  the situation is  far from being settled  down. Depending on the kind of mass tracer used (eg. stars, HI or stellar streams)  the shape of the   MW DM halo ranges from spherical, prolate,  oblate or  triaxial.  
This issue is expected to be resolved by  upcoming  larger  scale  surveys of the phase space distribution of stars in the MW, like GAIA, from which the gravitational potential can be recovered, and thereby also the distribution of DM. %

\subsection{The LG as a Dark Matter Laboratory: WDM vs CDM }
\label{sec:lgwdm}

\begin{figure*}
\includegraphics[width=16cm]{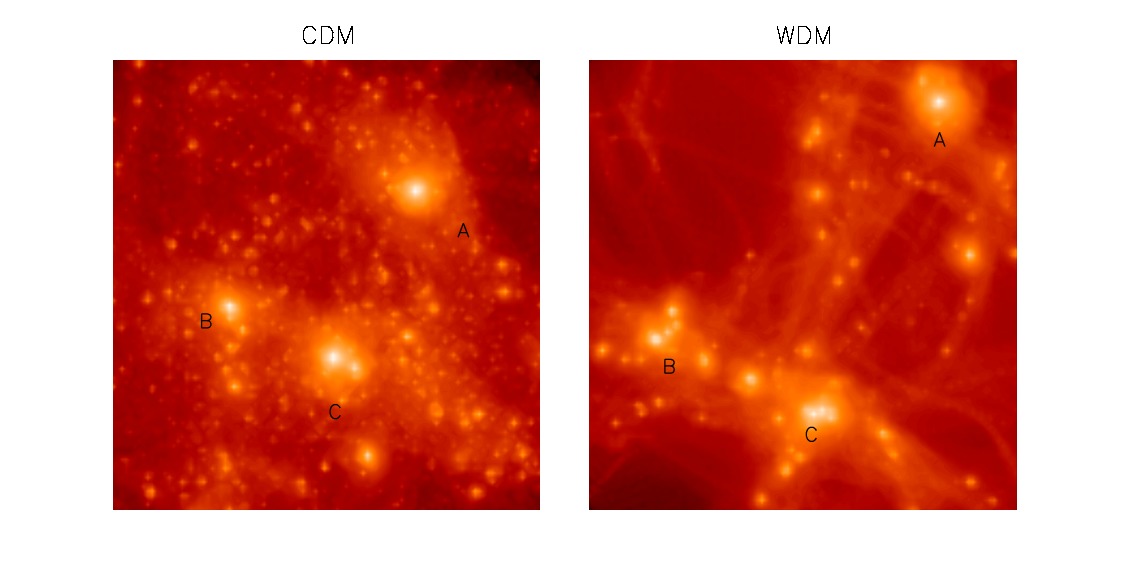}
\caption{ A comparison of  the dark matter distribution of the Local Group object in two different dark matter  scenarios, the standard CDM (CLUES simulation LG64-3)  and  in a WDM  scenario, assuming that the dark matter particles have a mass of 1 keV, (CLUES simulation  LG64-3-WDM)  (Figure taken from  \citet{wdmlgrev})}
\label{fig:wdmcomp}
\end{figure*}
As we have mentioned above, our Galaxy is the only place we can measure  dark matter structures at the  shortest scales. 
Therefore, we can use the observational information on the satellite
population of  MW and M31 to compare with the  predictions of simulations of MW-type halo formation  in different dark matter scenarios.  We have also mentioned the problem  the standard {\lcdm} model   has to explain the number of satellite galaxies  in the LG. High resolution N-body simulations predict  an order of magnitude more subhalos capable of hosting visible satellites than are detected in the LG.  The usual explanation to this discrepancy within the {\lcdm} model  is to resort to the existence of  biases  between the  satellite galaxies and dark matter  sub halos hosting them. The baryonic feedback processes can  efficiently suppress galaxy formation  inside them. Nevertheless, there is not only an inconsistency in the number of  galaxy satellites with respect to halos, but also about the kinematics of the observed dwarf spheroidals (dSphs) in  the MW.  The velocity profiles of
the most massive subhaloes found in high-resolution simulations of a
typical  $10^{12}$  \Msun   dark matter halo   \citep{bolyan2011} show
a  large peak circular velocity (assuming than dark matter follows an
NFW profile).  They are too dense to host  any of the bright
($L_V>10^5 {\rm L}_\odot$)  MW satellites. Moreover, they should
produce larger annihilation fluxes than the current detection limits
on dwarf galaxies set  by the FERMI satellite.   Possible solutions to
this puzzle have been explored within the {\lcdm} model by
\citet{bolyan2012}. None of them seems to be very convincing.  The
most plausible one was based on recent simulations by
\citet{governato2012} in which  they showed than the dark matter
density profile can become flatter (less concentrated)  if a large
amount of energy from supernovas   is able to blow large amounts of
gas from the centre of a small subhalo.  Nevertheless, the small
stellar content of the MW dSphs  indicates that this mechanism  is not
effective  in this case.   We have also shed some light into this interesting discussion using the CLUES simulations. \citet{2013MNRAS.431.1220D} have analysed the density profiles of dark matter in subhalos of the  LG64-3 and LG64-5 DM and SPH CLUES  simulations (see Table \ref{tab:resimu}) and concluded that the Einasto profile (eq. \ref{eq:einasto})  give a good fit  both  for subhalos in the DM and SPH simulations.

   Our conclusions are slightly different than in \citet{bolyan2011},
   which were based on assuming NFW density profiles.  We find that an
   Einasto profile with shape parameter $1.6 \le n_e\le 5.3$ provides
   an accurate matching between simulations and observations,
   alleviating the ”massive failure” problem first addressed in
   \citet{bolyan2011}.  However, in a follow-up paper by the same
   authors \citep{bolyan2012},  the direct particle data from the Aquarius simulations were used with no  appeal  to a specific fit  of the profiles and they confirmed  their previous results. 
   Nevertheless, as shown in the previous reference and, more
   recently,  by  \citet{vera2013},  a very good agreement with
   observation is attained  considering  that  the mass of the MW is
   smaller than $10^{12} $ \Msun. In our  WMAP3 CLUES  simulations we have
   slightly lower masses for MW and M31 ( $5.5 - 7.5 \times 10^{11}$
   \Msun).   This would   also reduce the probability of  the MW  hosting two satellites as bright  as the LMC and SMC.   
 Our  Galaxy  is in fact a rare one.   Different studies using the
 Sloan Digital Sky Survey (SDSS)  have concluded  that only  $\sim
 3.5$\% of the MW-like  galaxies  have  two satellites as big as  the
 Magellanic Clouds \citep[e.g.][]{Tollerud2011}. On the other hand, a low  mass MW  would be in conflict  with results from abundance-matching  
 based relations  between halo mass and stellar mass  \citep[see
 e.g.][]{bhz13},  direct measurements of   the MW's  virial mass  from
 kinematic studies  of  its satellites \citep{bk13}, or   the  recent
 measurement of the MW's  escape velocity  from  the RAVE survey  \citep{rave13}.  

Therefore,  due to the problems than {\lcdm } faces  to explain the
satellite population  in the LG,  the view has turned into  alternative models of dark matter in which either the  short scale power is suppressed because of  free streaming, assuming that the mass of the DM particles is in the keV regime (WDM) or   the dark matter is self-interacting,   producing  substructures with cored profiles  that can then easily be destroyed   \citep[e.g., ][]{vogelsberger2012,rocha12}. 

WDM  is an attractive alternative to CDM. The power spectrum of DM
fluctuations  has a sharp cutoff at short scales  due to  the effect
of the  thermal velocities of  the DM particles when  they became non relativistic. For the common assumption  of WIMPS as  DM candidate,  their masses are  in the GeV scales and thus, the relic thermal velocities of those particles are very small during the epoch of  structure formation.  If instead a keV mass particle DM candidate is considered,  as the sterile neutrino \citep[e.g.][]{asaka2005}, the thermal velocities are sufficiently large to erase, due to free streaming,  all  perturbations below   galactic scales.  In Fig \ref{fig:CDM_WDM_PK} we show a comparison between the  CDM and WDM power spectra for  the case of a  $m_{WDM} = 1 $ and $3$ keV.   On scales much  larger than the cutoff, structure formation proceeds very similar in both models, except for the  case that there is a delay in the formation of WDM  structures with respect to CDM.   This translates into smaller fluctuations at high redshift  than can be constrained  by comparing  with  astronomical observations of the Lyman $\alpha$ forest \citep[e.g.][]{viel2013}.  

WDM halos  of a Milky Way type  size have been recently simulated with high resolution  both with  collisionless N-body simulations  \citep[e.g.,][]{lovell2013}  as well as with  radiative  gasdynamical simulations  \citep[e.g.][]{maccio2013}).  In our CLUES project we have  also done the same experiments but focusing on the formation of the LG as a system of 3 main galaxies (LG64-3-WDM and LG64-3 CDM simulations of Table \ref{tab:resimu}). A comparison of the dark matter distribution  of the LG group in the two models, CDM and WDM (1 keV), at $z=0$ can be seen  in Fig. \ref{fig:wdmcomp}.   A detailed comparison of the internal properties of the main halos in these two models has been published recently  in \citet{wdmlgrev} and we refer the reader to this publication for further  information. Here we just want to  report a summary of our findings. As can be clearly seen in Fig \ref{fig:wdmcomp}, the  two LG groups are in a different stage of evolution. The CDM LG  is collapsing and more compact,   the WDM LG  is dynamically  younger, more diffuse and  is still expanding.  This delay  in the evolution  implies that WDM halos  are smaller than their CDM counterparts  at z=0.  They also show  lower baryon fractions  in their inner parts, where baryons dominate,  as compared  with  CDM.   The cutoff in the power spectrum also affects the baryonic processes in non trivial way: from the  star formation rates, to the  bulge/disk ratios  to colours of satellites. In general, the WDM  dwarf satellites tend to be less  gas rich and less concentrated.  We also find marginal evidence of a thickening of the disk gas in one of  our WDM galaxies as compared with their CDM counterpart, but this cannot be extrapolated as a general behaviour of  disk formation in WDM.  The  model used  in the CLUES simulations (1 keV WDM particle mass)  is  on  the  low side  of the allowed values  imposed by high-redshift Lyman $\alpha$  constrains ($2-3$ keV).  For higher masses of  WDM particles,  ($m_{WDM} >2 $ keV) the effects  on disk formation  have a minor impact \citep{maccio2013}. 

So, we  have seen how the LG can be considered as a  dark matter laboratory, not only because  it is our place in the universe in which we can try to  detect the nature of  dark matter (either by direct experiments on Earth or from indirect detection of their annihilation/disintegration  remnants), but also by comparing  astronomical observations of LG galaxies with the predictions of numerical simulations in different dark matter scenarios. We will continue with this issue in  section \ref{lu}  in which we will go a step forward in scale and will show how the observations of the Local Universe  can also help us in constraining the nature of dark matter.

\section{Formation of the Local Group and Local Group - like Objects}
\label{subset:LG}

\subsection{The Mass Aggregation History}
\label{subset:MAH}

The issue of how typical is the LG compared with similar objects in the universe is arguably one of the most  interesting  questions that can be asked within the framework of Near-Field Cosmology. 
Or, rephrasing it, is the LG typical enough that one can learn from it about the universe at large, thereby making the study of the near field a subfield of cosmology. 
\citet{2011MNRAS.417.1434F} have opted to
address the following aspect of the question. Namely, given the present
epoch dynamical constraints on the LG, to what extent the mass
accumulation history (MAH) of the LG is typical?
This   has been
tested by constructing ensembles of LG-like objects in a suite
of 3 different  CLUES LG64-5  DM   simulations (Table \ref{tab:resimu})  and the unconstrained
BOLSHOI simulation \citep{bolshoi}, which is used to provide the framework within which the question of 'how typical' is asked.

A numerical study of the LG should start with identifying the
main relevant observational features of the LG which will serve
as the basis for constructing an ensemble of LG-like objects.
\citet{2011MNRAS.417.1434F} used  DM-only simulations  to construct such an ensemble
and therefore considered only the dynamical properties of the LG and its
environment. The set of criteria used  in these studies includes the mass of the two main halos of the LG-like objects, their isolation, proximity to a Virgo-like halo, the distance between the two and the fact that they approaching one another  \citep[for a more detailed description see][]{2011MNRAS.417.1434F}.

Three different ensembles have been constructed out of the constrained simulations (combined) and separately out of the BOLSHOI simulation. 
The ensemble of individual halos consists of all haloes in the mass range $(0.5-5)  \times  10^{12}\ \hmpc$. 
The ensemble of pairs, where two halos, $H_{A}$ and
$H_{B}$,
from the {\it Individuals} sample are considered a pair if and
only if
halo $H_{B}$ is the closest halo to $H_{A}$ and
vice versa. Furthermore, with respect to each halo in the pair
there
cannot be any halo more massive than $5.0>10^{12}$\hMsun  closer
than
its companion. The sample of isolated pairs is of all pairs
which obey also the  environmental criteria for a
LG-like object. Hence this is the ensemble of LG-like objects.
In addition, the select group of the 3 simulated LGs of each one
of the constrained simulation forms a sample by itself.

\begin{figure}[ht]
\begin{center}
\includegraphics[scale=0.50]{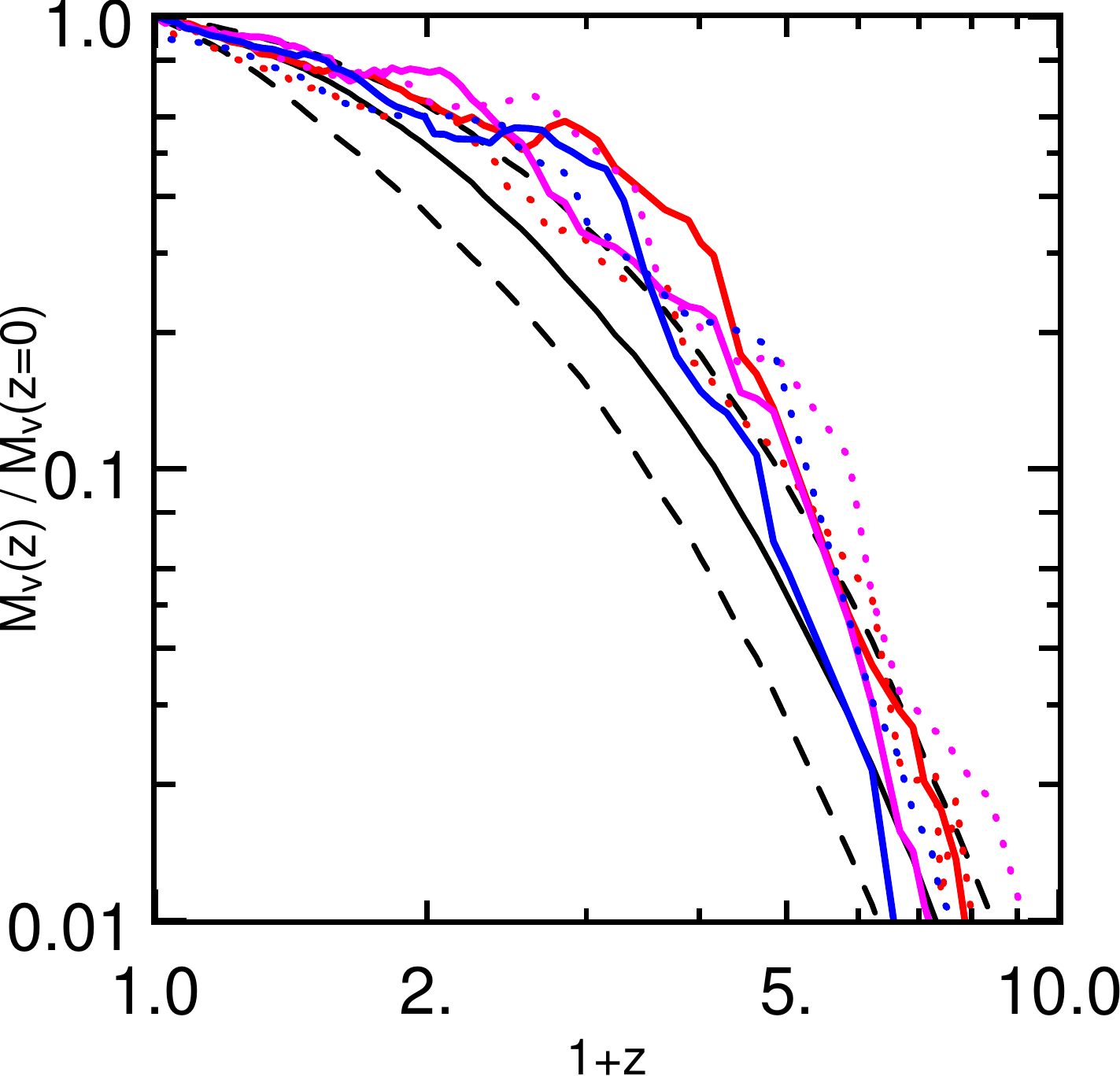}
\end{center}
\caption{  Mass assembly histories of LG halos in the LG64-5 CLUES
  simulation as a function of redshift. The solid black line shows the median
  MAH for all halos in the CLUES simulations within the mass range $5.0\times
  10^{11}\hMsun< M_{h}<5.0\times 10^{12}\hMsun$, the dashed lines show  the
  first and third quartiles. Also plotted as colour lines are the MAHs for the
  MW (dotted) and M31 halos (continuous)  in the three constrained
  simulations. The assembly history for the LG halos is systematically
  located over the median values as sign of early assembly with
  respect to all halos in the same mass range. (Figure taken from \citet{2011MNRAS.417.1434F}.)}
\label{fig:mah}
\end{figure}

The MAH of the selected halos is defined by three characteristic times, extracted from the merger trees of the halos.
The times, measured as look-back time in Gyr, are:  
a) {\bf Last major merger time} ($\tau_M$), defined as the time
    when the last FOF halo    interaction with ratio 1:10 starts. This
    limit is considered to be the mass ratio     below which the
    merger contribution to the bulges is $< 5\%$-$10\%$
    \citep{2010ApJ...715..202H}.
b). {\bf Formation time}  ($\tau_F$) marks the time when the main branch in the tree
    reached half of the halo mass at $z=0$. This signals the epoch when half of the
    galaxy mass (gas and stars) is already in place in a single collapsed object.
c). {\bf Assembly time} ($\tau_A$): defined as the time when the
    mass in progenitors more massive than M$_f=10^{10}$\hMsun is half
    of the halo mass at $z=0$. This time is related to the epoch of
    stellar component assembly, as    the total stellar mass depends
    on the integrated history of     all progenitors
    \citep{2006MNRAS.372..933N,2008MNRAS.389.1419L}. The exact
    value   depends  on $M_f$, and the specific value selected in this
    work was used to allow the comparison of assembly times against
    the results of the BOLSHOI simulation which has a lower mass
    resolution.

The MAH   of the three simulated LGs, drawn from the CLUES zoom simulations,  is presented 
  by Fig. \ref{fig:mah}, which shows the actual MAH of these objects.
The figure shows also the median and the   first and third quartiles
 of the  MAH for all halos in the CLUES simulations within the mass range $5.0\times
  10^{11}\hMsun< M_{h}<5.0\times 10^{12}\hMsun$. 

\begin{figure}
\begin{center}
  \includegraphics[scale=0.40]{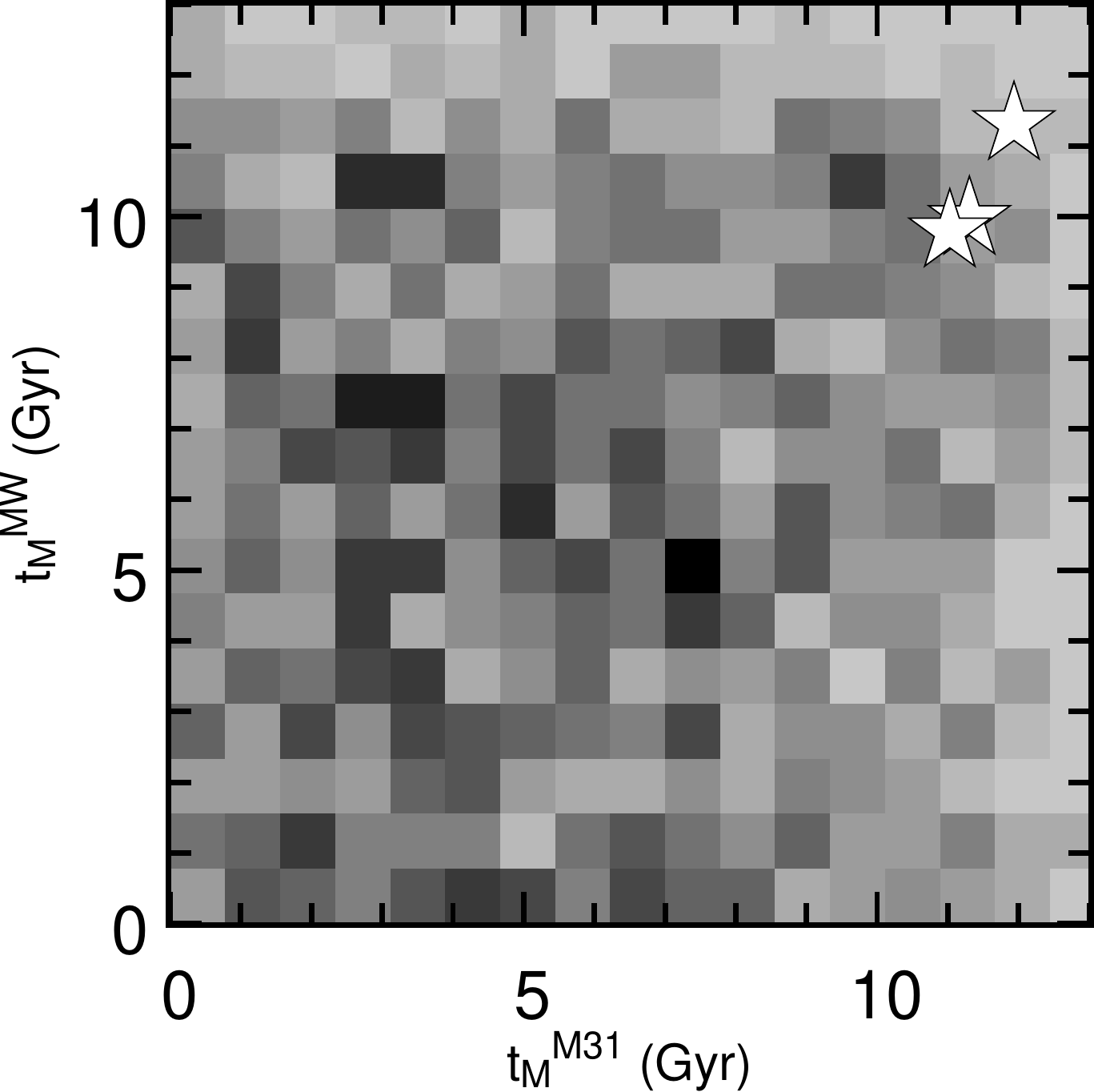}\hspace{0.5cm}
\includegraphics[scale=0.40]{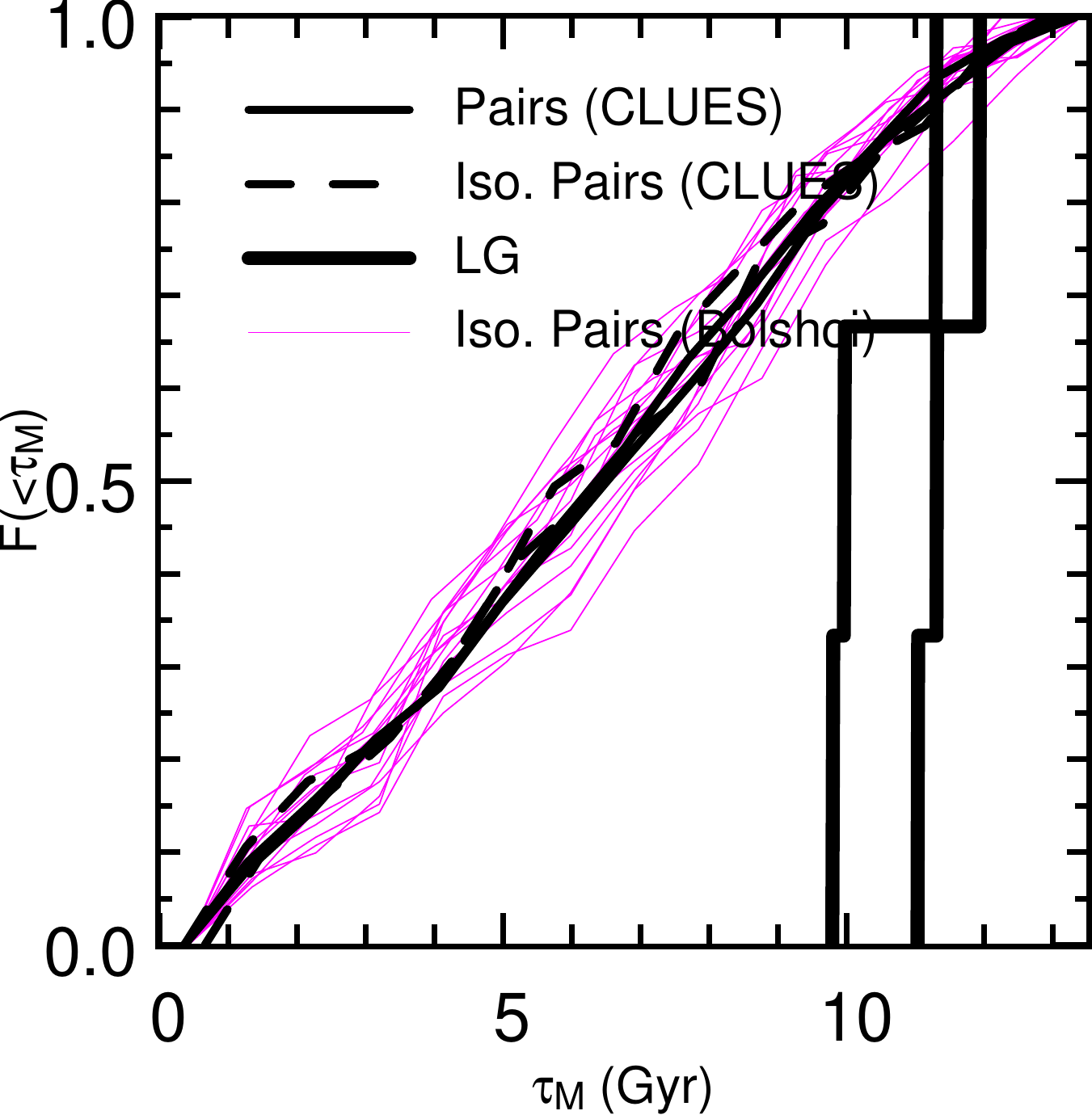}
  \includegraphics[scale=0.40]{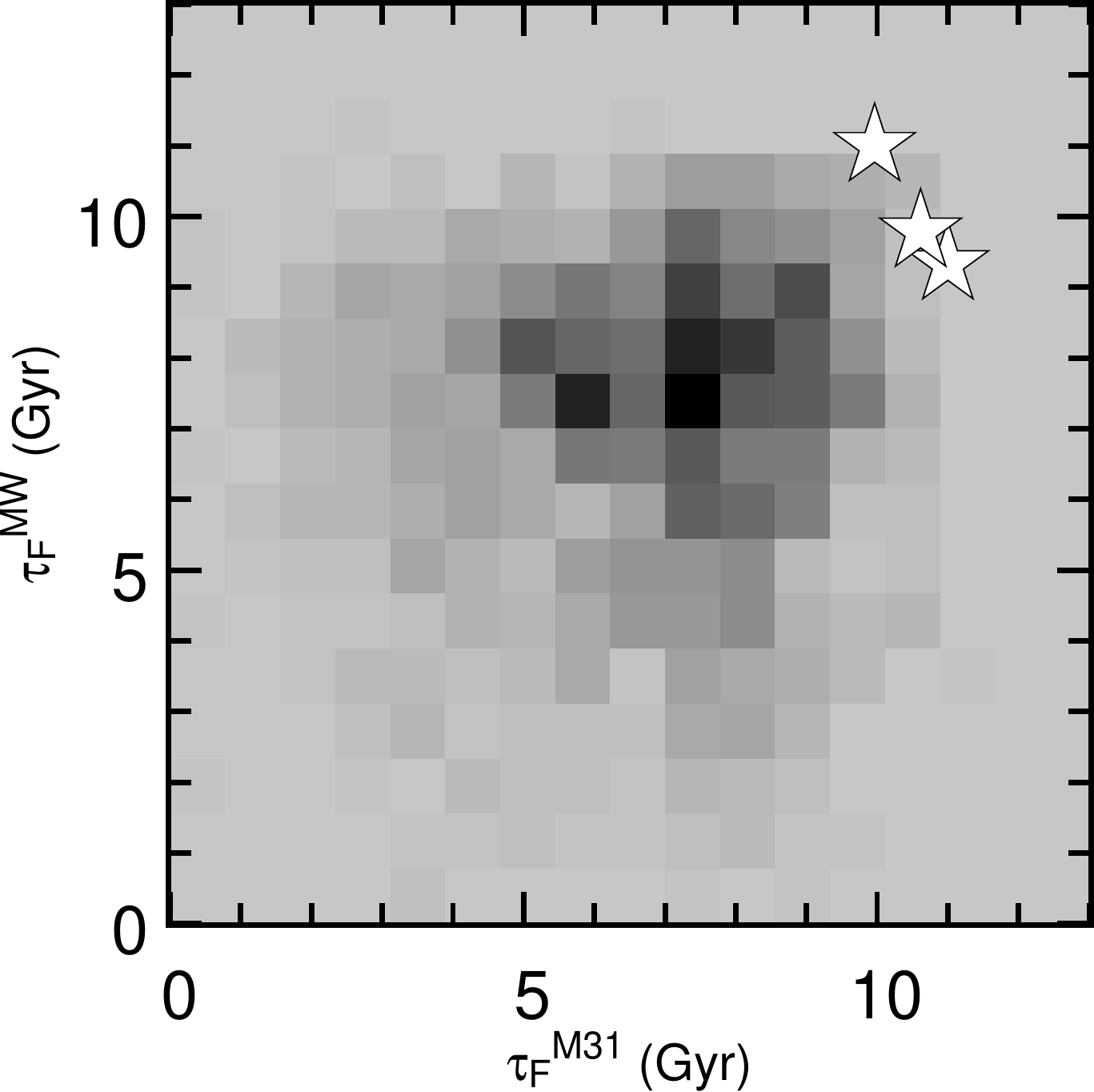}\hspace{0.5cm}
\includegraphics[scale=0.40]{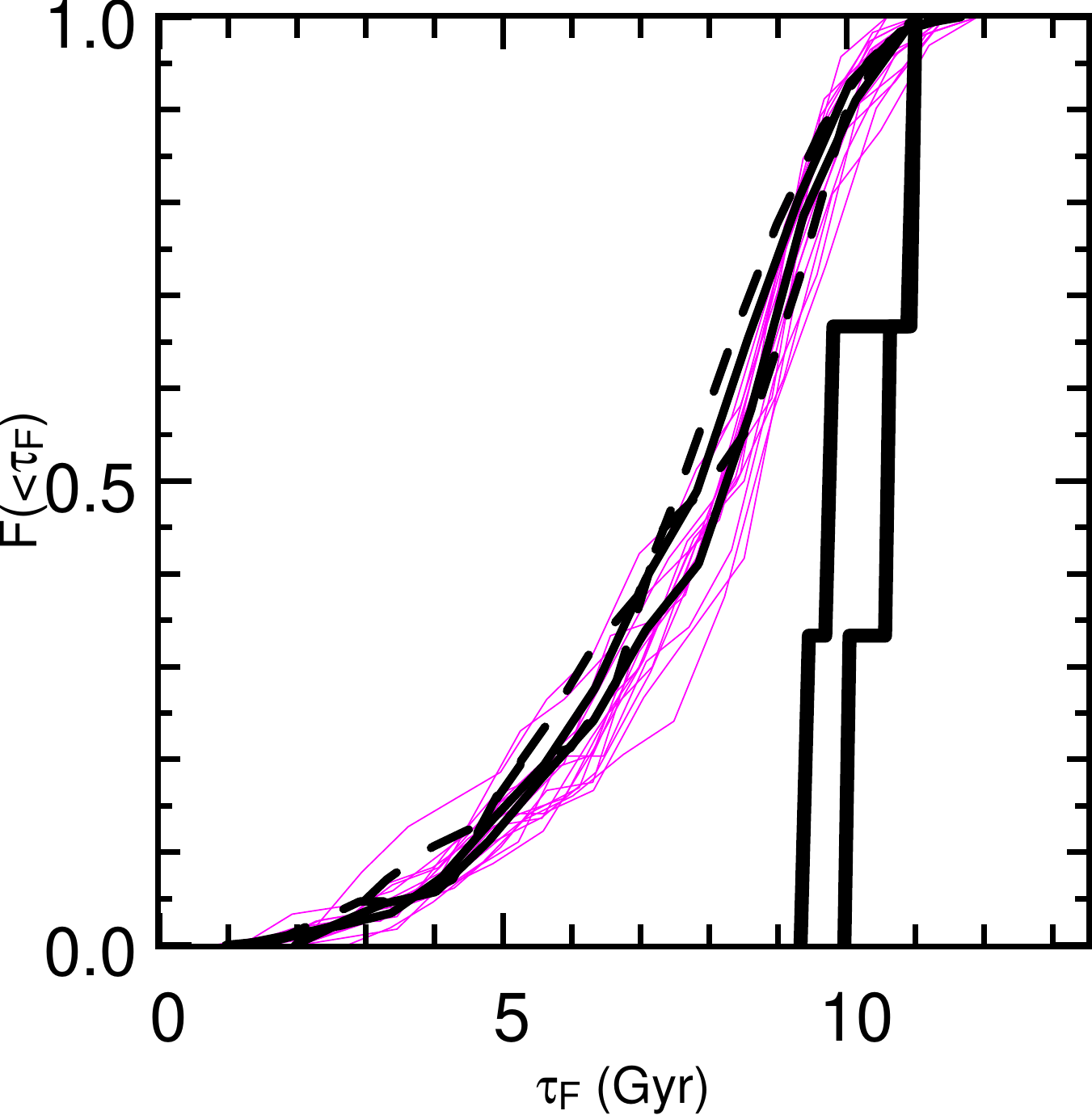}
\includegraphics[scale=0.40]{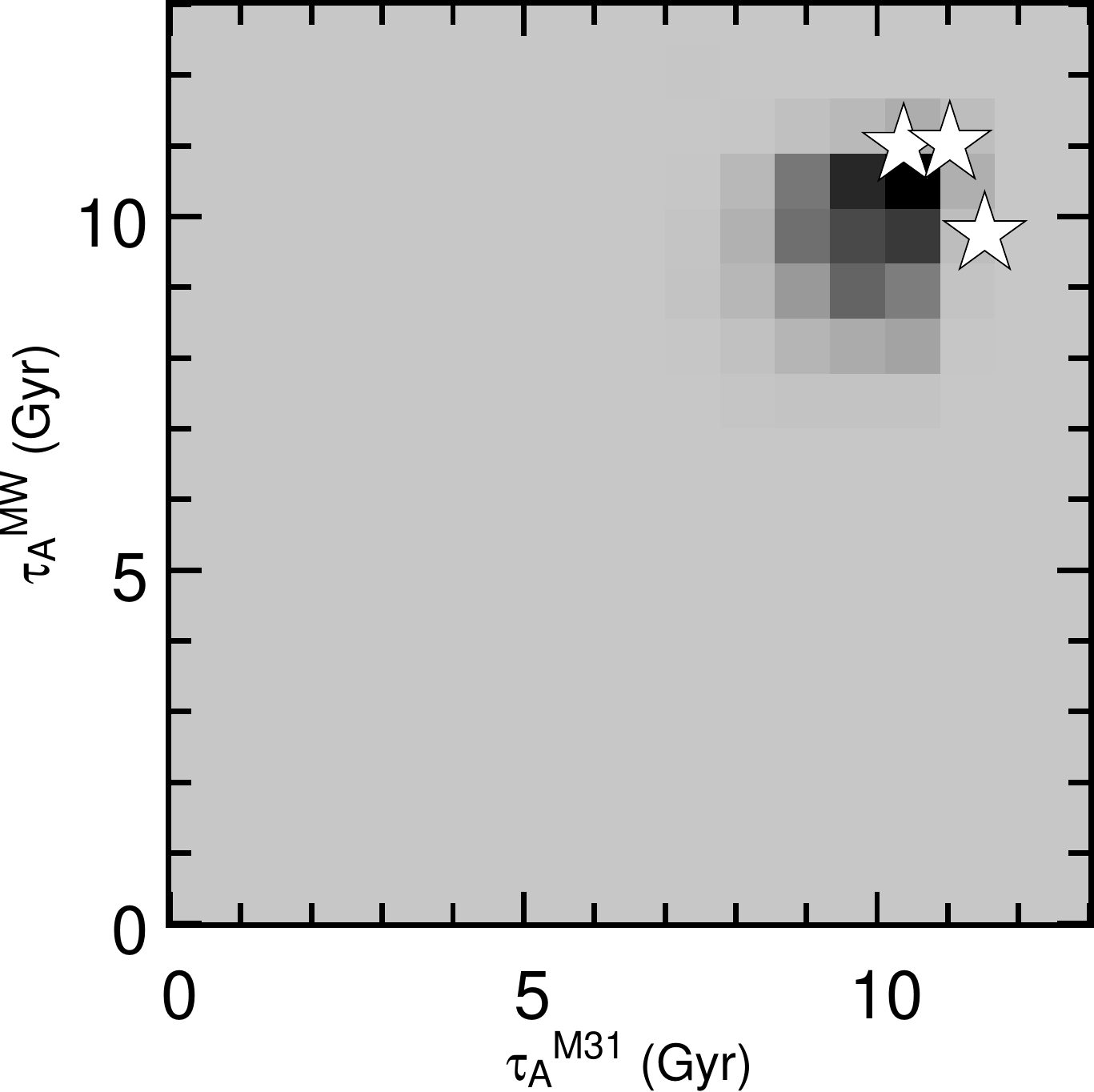}  \hspace{0.5cm}
\includegraphics[scale=0.40]{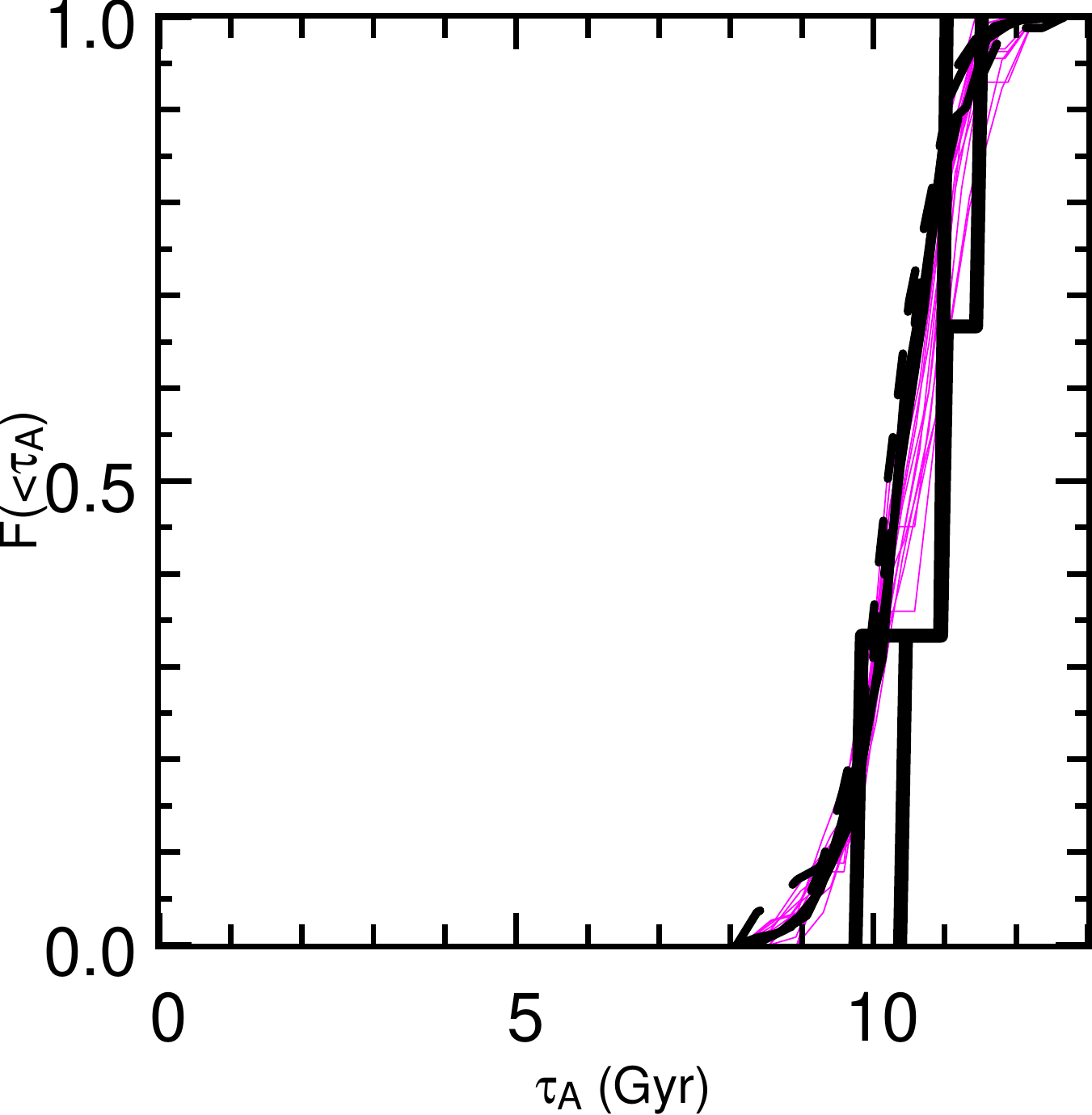}
\end{center}
\caption{{\it  Left column}. Joint distributions of three different times
  (last major merger, formation and assembly) describing the mass aggregation histories.
 Each point in the plane  represents a pair MW-M31 with histories described by the time values at that
 point. Levels in shading coding indicate the number of halo pairs in the
 BOLSHOI simulations in that parameter range. The stars mark the location of the three
 LG pairs from the constrained simulations. 
{\it Right column} Integrated probability of these three different
 times. The continuous black lines represent the results for the {\it
   Pairs} sample in the CLUES simulations. The {\it Isolated Pairs} sample from CLUES
 is represented by the thick dashed lines. The results from the {\it
   Isolated Pairs} samples in eight
 sub-volumes of the BOLSHOI simulation are represented by the thin continuous
 grey lines. The thick continuous lines represent the results for the
 {\it LG} sample.
 (Figure taken from \citet{2011MNRAS.417.1434F}.)
\label{fig:surface}}
\end{figure}

Fig. \ref{fig:surface} provides the essence of the analysis. It shows the joint distribution of the three MAH times of the different samples. 
 Two basic facts immediately emerge here. 
One, the 3 simulated LGs show a remarkable clustering in the joint phase space of the MW and M31 MAH characteristic times. 
This fact is not trivially expected from the selection and construction of the simulated LGs. 
The other is the distributions from the {\it Pairs} and {\it Isolated Pairs}
 control samples are basically indistinguishable. In other words,
 detailed selection criteria for halo pairs, based on isolation only,
 do not narrow   significantly the range of dark matter halo
 assembly properties. 

The main finding of \citet{2011MNRAS.417.1434F} is that the three LGs share a similar MAH with
formation and last major merger epochs placed on average $\approx 10-12$
Gyr ago. Roughly $12\%$ to $17\%$ of the halos in the mass range
$10^{11}\hMsun <M_{h} < 5\times 10^{12}\hMsun$ have a similar MAH. In
a set of pairs of halos with similar characteristics as the LG, a
fraction of $1\%$ to $3\%$ share similar formation properties as both halos in the
simulated LG. An unsolved question posed by our results is the
dynamical origin of the MAH of the LGs. The isolation criteria
commonly used to define  LG-like halos do not reproduce such a quiet
MAH, nor does a further constraint that the LG resides in a low
density environment.                                                              
The quiet MAH of the LGs provides a favourable environment for the
formation of disk galaxies like the MW  and M31. The timing for
the beginning of the last major merger in the Milky Way dark matter
halo matches with the gas rich merger origin for the thick component
in the galactic disk.  Our results support the view that the specific
large scale environment around the Local Group has to be explicitly considered
if one wishes to use near field observations as validity test for $\Lambda$CDM.  

\subsection{The kinematics of the Local Group in a cosmological context}
 \label{subset:kinematics}

Recent observations constrained the tangential velocity of M31
with respect to the MW  to be $v_{\rm M31,tan}<34.4
\kms$ and the radial velocity to be in the range $v_{\rm
M31,rad}=-109\pm 4.4 \kms$ \citep{2012ApJ...753....9V}.
This has motivated \citet{2013ApJ...767L...5F} to 
revisit the question 
of how typical is the LG with respect to new kinematical data,
using the   simulations and selection of objects of \citet{2011MNRAS.417.1434F}. %
The following main findings have emerged.
The most probable values for the tangential and radial
velocities in these pairs are $v_{\rm rad, \Lambda CDM}=-60\pm
15 \kms$ and $v_{\rm tan, \Lambda CDM}=50\pm 5 \kms$.
Using the same absolute values for the uncertainty in the
observed velocity components, halos within the preferred
$\Lambda$CDM values are found to be five times more common than
pairs compatible with the observational constraint. The
qualitative nature of these results is still valid after a
narrower selection on separation and total pair mass.
Additionally, pairs with a fraction of tangential to radial
velocity $f_{\rm t}<0.32$ (similar to observations) represent
$8\%$ of the total sample of LG-like pairs. Making a tighter
selection to match the observational constraints on the
separation and total mass results in zero pairs compatible with
observations.

Approximating the LG as two point masses the above mentioned
results are expressed in terms of the orbital angular momentum
$l_{\rm orb}$ and the mechanical energy $e_{\rm tot}$ per unit
of reduced mass. Uncertainties in the tangential velocity, the
square of the norm of the velocity and the total mass in the LG
lead to poorer constraints on the number of simulated pairs that
are consistent with the observations. Nevertheless, in the case
of the LG-pair sample that also fulfills the separation and
total mass criteria there is a slight tension between simulation
and observation.

The kinematics of the three simulated LGs is dominated by radial
velocities. However their velocity components differ from the
observational constraints while their mechanical energy and
orbital angular momentum are in broad concordance with
observations. There is only one pair that fulfills all the
separation, total mass constraints and matches the most probable
value for the dimensionless spin parameter $\lambda$ inferred
from observations.
LG-like pairs in $\Lambda$CDM show preferred values for their
relative velocities, angular momentum and total mechanical
energy. The values for the orbital angular momentum and energy,
merged into the $\lambda$ spin parameter, are in mild
disagreement with the observational constraints. However, there
is a strong tension with the precise values for the radial and
tangential velocities.
This leads to an interesting observation, assuming, in a very
rough approximation, that the mechanical energy and total
orbital angular momentum of the LG are conserved, as in the
classical two-body problem. Then the consistency of the energy
and orbital angular momentum of the simulated LGs with the
observed one, and the inconsistency with respect to the actual
values of the radial and tangential velocities, implies that the
observed LG started from initial conditions that are not
captured by the simulated LGs. The unique phase of the LG on its
orbit is not reproduced by the simulations reported here.

\section{Weighting  the Local Group}
\label{weight}

The estimation of the mass, be it the total, DM or the baryonic
mass, of cosmological objects such as galaxies, haloes, groups
and clusters of galaxies, is a very challenging task. In many
ways the problem is an ill defined one. The cosmological
structures mentioned here are not isolated   and  
 are not well %
separated from the continuous mass
distribution in the universe. This is also valid in the
theoretical domain, where mass estimation is unaffected by
observational limitations. Yet, algorithms are defined and halo
finders are devised so as to provide mass estimators which
conform with theoretical understanding and observational
feasibility  \citep[see][and references there in]{2013MNRAS.428.2039K,2013MNRAS.435.1618K}.

The CLUES  simulations provide a unique numerical laboratory for testing mass
estimators designed to assess the mass of very local objects of
interest, such as the the MW and M31 galaxies, nearby satellites
and dwarf galaxies and the LG as an object by itself.

\subsection{Mass estimation by the Timing Argument}
\label{subset:TA}

It has been known for a  long time that the total mass of
the LG can be estimated by assuming it to be a two-body system,
i.e. an isolated system made of two point mass particles whose
individual masses do not change with time. A further assumed
simplification is that  the two galaxies
formed at the time of the Big Bang at zero distance, hence their
orbital angular momentum is zero.
The so-called ``mass estimation by the timing argument'' is
based on the fact that in the above simplified model the age of
the universe, the distance between the MW and M31 and their
relative velocity determine  the total mass of the two objects,
and it  can be easily calculated  \citep{kahn-woltej,1981Obs...101..111L}. 
 A naive examination of the model would cast serious doubts on the
ability of the model to produce a useful prediction of the mass.
The LG is not an isolated system, the orbital angular momentum
of the two galaxies is not necessarily zero, the two galaxies
and their halos are not point-like particles, and their masses
certainly change over their dynamical evolution. It is therefore
not surprising that the mass estimation by the timing argument
(TA) has not played a prominent role in the effort to determine
the mass of the LG.
\citet{1991ApJ...376....1K} and \citet{2008MNRAS.384.1459L} took
the steps towards calibrating the model and make it into a
quantitative tool. The calibration consists of identifying
LG-like pairs of halos in large scale cosmological simulations
and applying the TA mass estimator to these objects.

\begin{table}[ht]
\begin{center}
\begin{tabular}{lccccc}
\hline
\hline

Simulation & \# of pairs & 25\% & 50\% & 75\% & $\frac{\Delta\eta}{\eta}$\tabularnewline
\hline
Box64-3 & 14 & 0.80 & 1.40 & 1.90 & 0.29\tabularnewline
\hline
Box64-5, A & 9 & 1.28 & 1.36 & 1.81 & 0.19\tabularnewline
\hline
Box64-5, B & 23 & 1.35 & 1.83 & 2.81 & 0.40\tabularnewline
\hline
Box64-5, C & 13 & 1.06 & 1.46 & 1.90 & 0.29\tabularnewline
\hline
Box64-5, all & 45 & 1.23 & 1.55 & 2.03 & 0.26\tabularnewline
\hline
L\&W & 16479 & 1.07 & 1.48 & 2.12 & 0.35\tabularnewline
\hline
\hline
\end{tabular}
\end{center}
\caption{The $\eta_{vir}$ percentile values for the Box64-3   the three Box64-5 CLUES  simulations, the distribution  for the three Box64-5  simulations combined. The  last row shows the results of \citet{2008MNRAS.384.1459L} and  are presented for reference.}
\label{table:eta-vir}
\end{table}

The CLUES  has been used for testing
the TA mass estimation\footnote{\texttt{Shalhevet
Bar-Asher, 2011, MSc. Thesis, Hebrew University, unpublished}}.
That study consists of analysing a suit of the  Box64-3  and three
different realizations of Box64-5  DM only  
simulations (see  Table \ref{tab:fullbox}), construction of an ensemble of LG-like objects, following
the selection criteria outlined in \S \ref{subset:MAH}. 
Each
one of the simulations harbours a ``good LG" at its centre, namely
a simulated objects which obeys all the selection criteria for
being considered as a numerical counterpart of the observed LG.
Here the mass by the TA is compared with the virial mass of the
LG, namely the sum of the virial masses of the two most massive
members of the LG-like object.
Table \ref{table:eta-vir} shows the number of LG-like objects in
each simulations, the median and the $25\%$ and $75\%$ quartiles
of the distribution of $\eta_{vir}$, the ratio of the virial
masse of the LG-like object to the mass estimated by the TA
($\eta_{vir}=M_{vir} / M_{TA}$). The table shows also the
results obtained by \citet{2008MNRAS.384.1459L} as reference.
One should note here that the later study is based on the
$\Lambda$CDM cosmology but with a different set of cosmological
parameters and a different selection of simulated LG-like
objects than those reported here.

The results presented here are consistent with the findings of
\citet{2008MNRAS.384.1459L}, with a median value of $\eta_{vir}$
of about $1.5$, but with a smaller scatter. It follows that the
TA systematically biases the estimated mass towards smaller than
the actual virial mass. Adopting the following parameters for
the LG, an infall velocity of $130 \kms $, a distance of $0.784\
$  Mpc and assuming the WMAP5 age of the universe of $13.75$ Gyr,
we find  $M_{TA}=5.3\times 10^{12}\ \Msun$.  
The calibration from TA mass estimation to the virial mass yields $M_{vir}=\big(8.2{^{+2.5}_{-1.6}}\big)\times 10^{12} \ \Msun$.
The error bars reflect the theoretical uncertainties in
the $M_{TA} \ - \ M_{vir}$ relation, which are much larger than
the observational uncertainties,  which  are therefore ignored
here.

\subsection{Mass estimation by the Timing Argument in the presence of dark energy}
\label{subset:TA-Lambda}

The classical TA model is formulated within a cosmological model that
ignores the presence of a dark energy (DE)  component \citep{kahn-woltej,1981Obs...101..111L}. Yet, a DE component changes the time behaviour of the background universe, and hence is expected to affect the TA model. This has been recently considered by  \citet{2013arXiv1308.0970P}. In Newtonian terms the DE is acting to provide an effective repulsive force, and therefore the modified TA model is expected to result in a higher estimated mass. Using the latest compilation of cosmological parameters of the PLANCK CMB experiment the modified TA model yields $M_{TA}=(4.73 \pm 1.03)\times 10^{12} \Msun$, an estimation which is 13\% higher than the original TA model
\citep{2013arXiv1308.0970P}. 
A calibration of the modified TA model  by the sample of LG-like objects (reported in \S \ref{subset:TA}) yields $M_{vir}/M_{TA}=1.04 \pm 0.16$. Applying it to  the TA   mass of the LG   results in an   estimated   virial mass of $M_{LG} = (4.92 \pm 1.08 (obs.) \pm 0.79(sys.)) \times 10^{12} \Msun$ 
\citep{2013arXiv1308.0970P}.

\subsection{Mass Estimators of the MW and M31 Galaxies }
\label{subset:ME}

Gas rotation curve data 
provide a reliable estimation of the mass distribution within the innermost few tens
of kiloparsec, of  both the MW and the M31 galaxies,  but estimation of the mass out to the virial  radius needs to rely on the kinematics of a tracer populations.
These tracers can either be globular clusters, halo stars or
satellite galaxies or a combination of these.
Earlier efforts in that direction include the mass of four MW  dwarf spheroidals (dSphs) satellites that  were constrained with
high precision 
by kinematic data sets \citep{Lokas09}.
Line-of-sight kinematic observations enable accurate mass
determinations at half-light radius for spherical galaxies such
as the MW dSphs \citep{Wolf10}: at both larger and smaller
radii, however, the mass estimation remains uncertain because of
the unknown velocity anisotropy.

The mass estimation of our own Galaxy, the MW, is about to be
revolutionized by the upcoming data
of the space mission GAIA  which will
provide
full six-dimensional phase-space information for all objects, in
the 
nearby   %
universe, brighter than $G\approx20$ mag.
GAIA's mission is to create the largest and most precise three
dimensional chart of the Milky Way by providing precise
astrometric data like positions, parallaxes, proper motions and
radial velocity measurements for about one billion stars in our
Galaxy and throughout the LG.

Anticipating the GAIA upcoming data and a situation at which the
position and proper motion data of satellite galaxies of the MW
will be available,  \citet{Watkins10} suggested a suit of
``scale-free projected mass estimators" to calculate the mass of
the MW. The estimators are based on simplifying assumptions such
as spherical symmetry and a scale-free density profile, which
constitute only a proxy to the actual MW. This motivated
\citet{2012MNRAS.423.1883D} to test \citet{Watkins10} mass
estimators against the  simulated MW and M31 of the CLUES LG64-5
high resolution zoom simulation.   It was
shown before that the considered  mass estimators work rather
well for isolated spherical host systems, and this was extended
by
\citet{2012MNRAS.423.1883D}
to examine their applicability to the MW and M31 haloes 
that form a binary system with a distinct satellite population.
Their analysis consists of the application of the mass
estimators using a number of sub-haloes similar to the number of
observed satellites of MW and M31, $N\sim30$, with the same mass
range and following the same observed radial distribution. It
has confirmed the notion that the scale-free estimators work
remarkably %
well -- even in our constrained simulation resembling a
realistic numerical model of the actual Local Group (as opposed
to isolated MW-type haloes considered in other works). It has
further validated that the accuracy increases when the full
phase-space information of the tracer objects is assumed. The
study has demonstrated, in the isotropic case, that no further
assumptions are required with respect to the host's density
profile: under the assumption that the satellites are tracking
the total gravitating mass the power-law index $\gamma$ derived
from the radial satellite distribution $N(<r)\propto
r^{3-\gamma}$ is directly related to the host's mass profile
$M(<r)\propto r^{1-\alpha}$ as $\alpha=\gamma-2$; it has been
shown that utilizing this relation for any given $\gamma$ will
lead to highly accurate mass estimations within our numerically
modelled LG. This is a fundamental point for observers and the
applicability of the scale-free mass estimators, respectively,
since the mass profile of the MW and M31 haloes is not a priori
known. Although future missions will improve the census of
satellite galaxies, it has been asserted that mass estimators of
the type studied by \citet{Watkins10} can already be safely
applied to the real MW and M31 system, and will acquire even
more importance with the forthcoming GAIA data.

\section{ Dark Matter distribution in the Local Universe: The Cosmic Web}
\label{cosmicweb}

The translation of the  vivid visual impression of the
cosmic web into a quantitative mathematical formalism poses an
intriguing challenge. There are two basic approaches to the
problem. One is observationally motivated, and it essentially
aims at defining the web from the point distribution presented
by the observed galaxy distribution
\citep{1999MNRAS.302..111L,2006MNRAS.366.1201N,2007A&A...474..315A,2008MNRAS.383.1655S}.
The other approach is motivated by the theoretical quest to
understand the emergence of the cosmic web and it is more
applicable to numerical simulations rather than observational
redshift surveys
\citep{2007MNRAS.375..489H,2009MNRAS.396.1815F}. The V-web web
finder, which follows  the later approach, is based on Clouds-in-Cells (CIC) interpolating the velocity shear tensor on a  grid, and calculating the number of the
velocity shear tensor eigenvalues above a certain threshold. The number of eigenvalues above a
threshold determines the web type of a given cell on the grid.
The V-web finder has two free parameters which
determine the web, the Gaussian smoothing of the CIC gridded
fields (density, velocity, etc.) and the threshold for the
eigenvalues of the (normalized by the Hubble constant) velocity
shear tensor
\citep{2012MNRAS.425.2049H,2012MNRAS.421L.137L,2013MNRAS.428.2489L}.

The BOX64-5 DM-only simulation (see Table \ref{tab:fullbox}) is used here as a
proxy to the actual nearby LSS. The simulation contains a
simulated LG-like at the centre of the computational box. The
analysis    focuses first on the DM distribution and the
cosmic web properties of the full computational box and then it
zooms to study the properties and structure of the web in the
immediate neighbourhood of the LG. A Gaussian kernel smoothing of
$R_s=0.25 \hmpc$ and a dimensionless threshold of 0.45 are used
to determine the V-web.

\begin{table}[ht]
\begin{center}
\begin{tabular}{lcccc}
\hline
\hline
web elements & volume filling fraction & mass filling fraction  \tabularnewline
\hline
voids                &  $0.68$ $(0.69)$          &   $0.13$ $(0.15)$     \tabularnewline
\hline
sheets               &  $0.27$ $(0.26)$          &   $0.36$ $(0.37)$     \tabularnewline
\hline
filaments           &  $0.046$  $(0.046)$          &   $0.34$ $(0.37)$     \tabularnewline
\hline
knots                  &  $0.0036$ $(0.0035)$          &   $0.17$  $(0.11)$     \tabularnewline
\hline
\hline
\end{tabular}
\caption{
 The volume and mass filling factors of the various web elements. The filling factors obtained without the multi-scale correction are given in the parentheses.
}
\label{tab:filling}
\end{center}
\end{table}

\begin{figure*}[ht]  
\begin{center}  
\includegraphics[width=6.5cm]{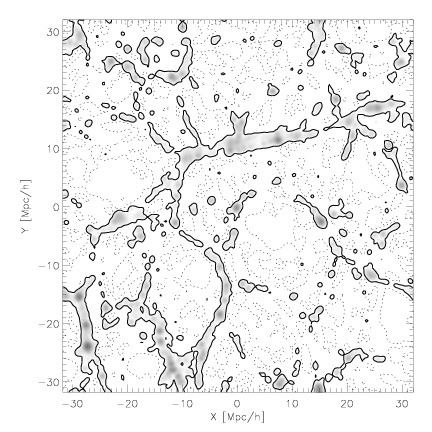} 
\includegraphics[width=6.5cm]{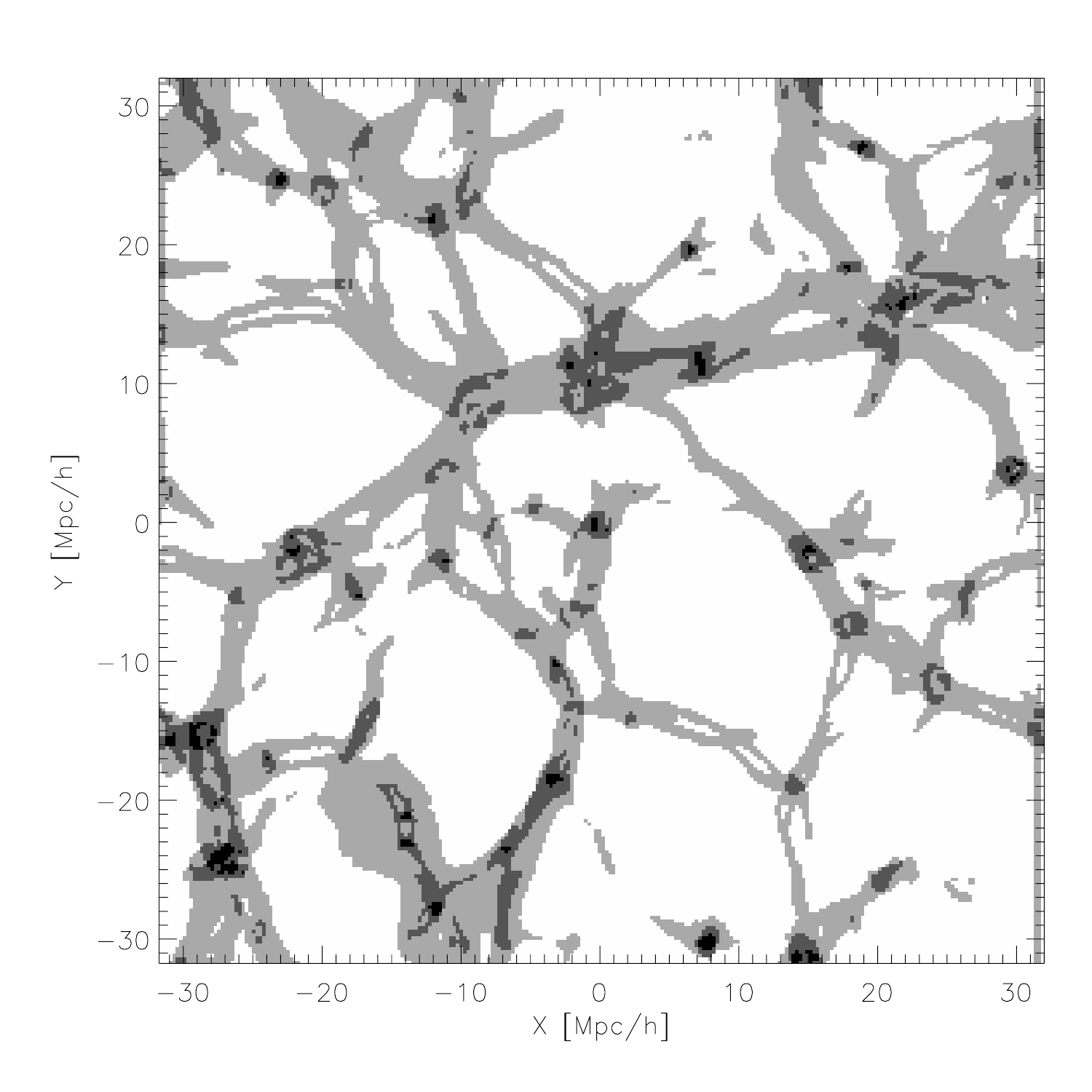}   
\caption{The normalized density   field and the cosmic web, 
based on the CIC density field of the full computational box (BOX64) spanned on a $256^3$ grid and Gaussian smoothed on the scale of $0.25 \hmpc$:
a.The density field presented by $\log\Delta$ (grey scale correspond to overdense  and dashed contours to under-dense regions.
 The solid contour represents the mean density. 
 (upper-left panel).
b. The velocity based cosmic web generated with  $\lambda{^V_{th}}=0.44$ 
made of voids (white), sheets (light grey), filaments (dark grey) and knots (black). Note that the map presents a planar cut through the cosmic web, hence sheets appear as long  filaments and filaments as isolated compact regions.
(Figure taken from \citet{2012MNRAS.425.2049H}.)
}  
\label{fig:box64-web} 
\end{center} 
\end{figure*}

\subsection{The Local Dark Matter Distribution and the Cosmic
Web}
\label{subsec:LocalCosmicWeb}

Fig. \ref{fig:box64-web} presents the normalized density,
$\Delta$, field and the cosmic web of the simulated
Supergalactic Plane, where $\Delta=\rho / \bar{\rho}$ and
$\bar{\rho}$ is the cosmological mean density. The left panel
shows the logarithm of the normalized density field and the
right one  exhibits the corresponding cosmic web. Both the
density field and web are based on the CIC gridded and smoothed
with a Gaussian kernel of $0.25\hmpc$ fields. 

Both panels of Fig. \ref{fig:box64-web} show a two dimensional
cut in a three dimensional computational box, hence the
filamentary-like looking structures that dominate the density
map and the cosmic web are actually sheets that are bisected by
the Supergalactic Plane. Most of the  actual filaments appear in these
two-dimensional maps as compacts knots. The LG is located within
a compact knot (black compact region in the web map), embedded
in a filament that runs perpendicular to the Supergalactic Plan.
The LG neighbourhood, up to a distance of
about $15 \hmpc$ is dynamically dominated by the Local Supercluster, a
structure that harbours the simulated Virgo and Ursa Major
clusters. The LG itself resides within an under-dense region,
characterized as a void by the V web finder, bisected by the
sheet, that contains the filament that contains the LG.

Table \ref{tab:filling} present the mass and volume filling
factors of the voids, sheets, filaments and knots of the full
computational box. The table shows that while most of the volume
of the box is taken by the voids, the sheets and filaments
contain about a third of the total mass of the box, with the
rest spreads almost evenly between the voids and knots. The
threshold which defines the V-web  has
been chosen so as to match the visual inspection of the
simulation (see \citet{2012MNRAS.425.2049H}). 
This threshold  roughly divides the mass
distribution into two halves, namely the mass filling factors of
voids and sheets combined, and of the filaments and knots
combined are roughly equal, $\approx 0.5$.

\begin{figure*}  
\begin{center} 
\includegraphics[scale=0.5]{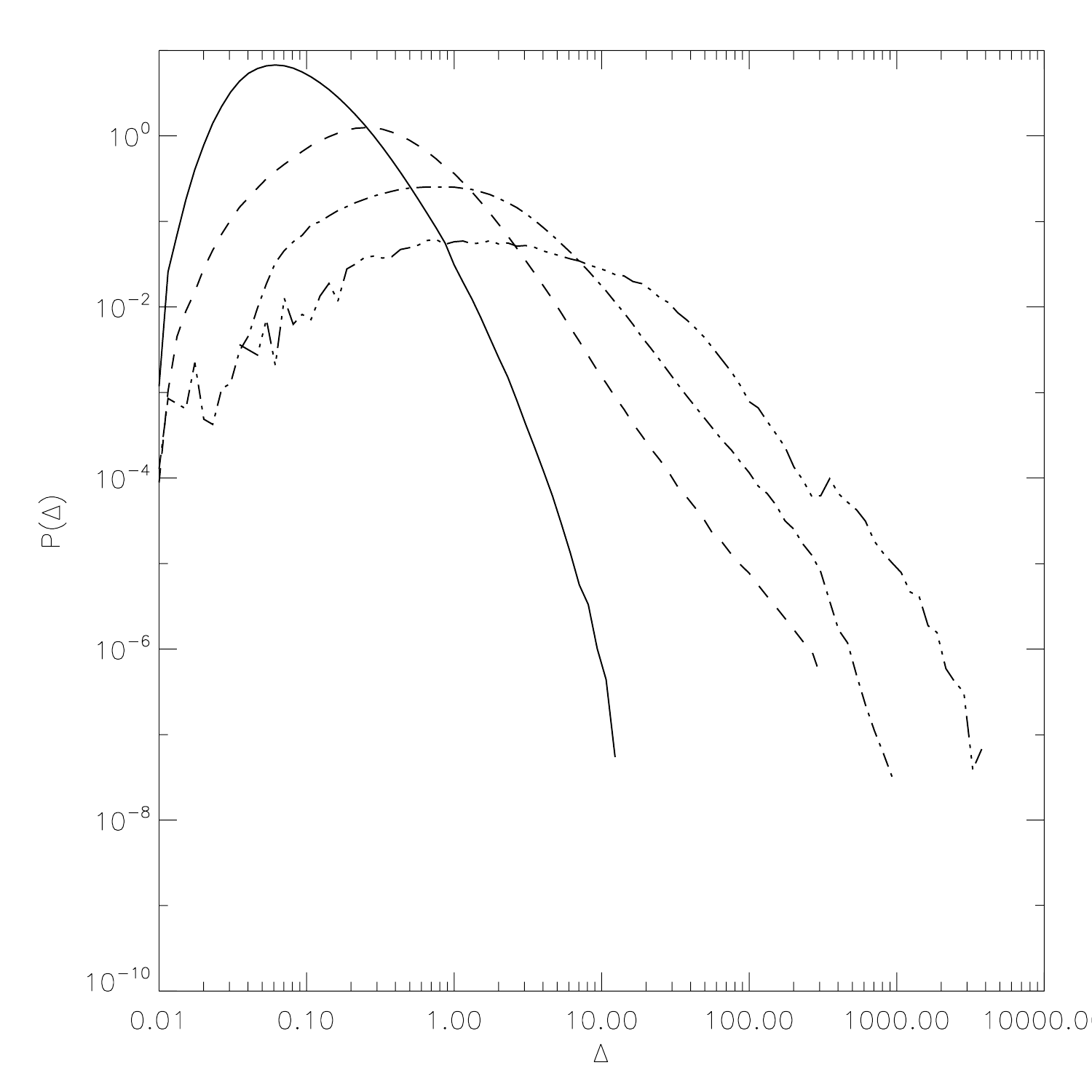} 
\caption{The  probability  distribution of grid cells  as a function of the fractional density,   $P(\Delta)$,  
 is plotted   for the various V-web elements, voids (full line), sheets (dashed), filaments (dot-dashed) and knots (dot-dot-dashed). 
 (Figure taken from \citet{2012MNRAS.425.2049H}.)
}
\label{fig:hist} 
\end{center} 
\end{figure*}

Fig. \ref{fig:hist} shows a histogram by CIC cells of the
normalized density field for the different web elements. The
histogram shows the relative abundance of CIC cells as a
function of their (normalized) density. It clearly shows that
although there is a strong correlation of the density of a CIC
cell with its web type there is no one-to-one correspondence
between the web classification and the density.

\subsection{The Local Group and the Cosmic Web}
\label{subset:LGCosmicWeb}

\begin{figure*}  
\begin{center} 

\includegraphics[width=6.60cm]{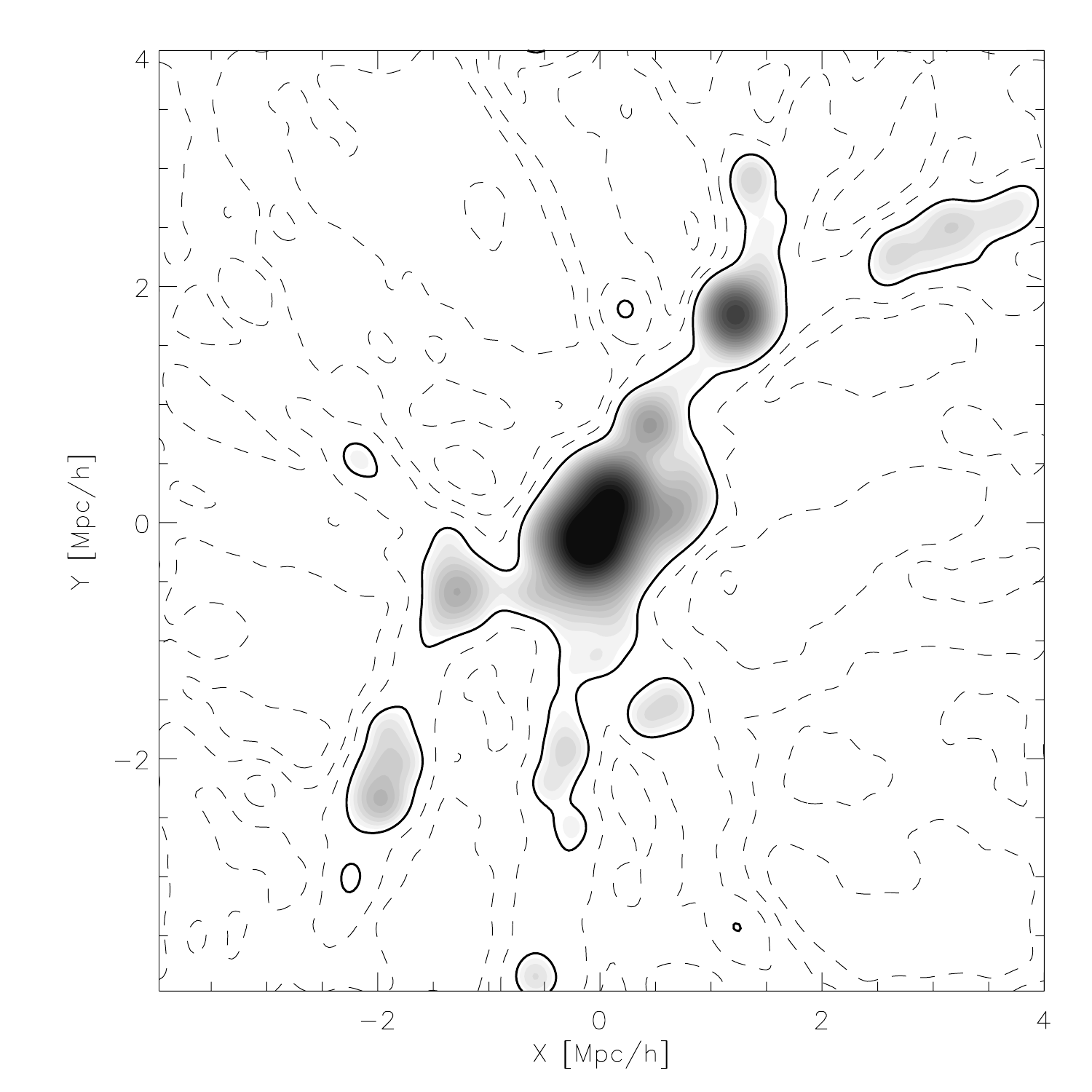}
\includegraphics[width=6.60cm]{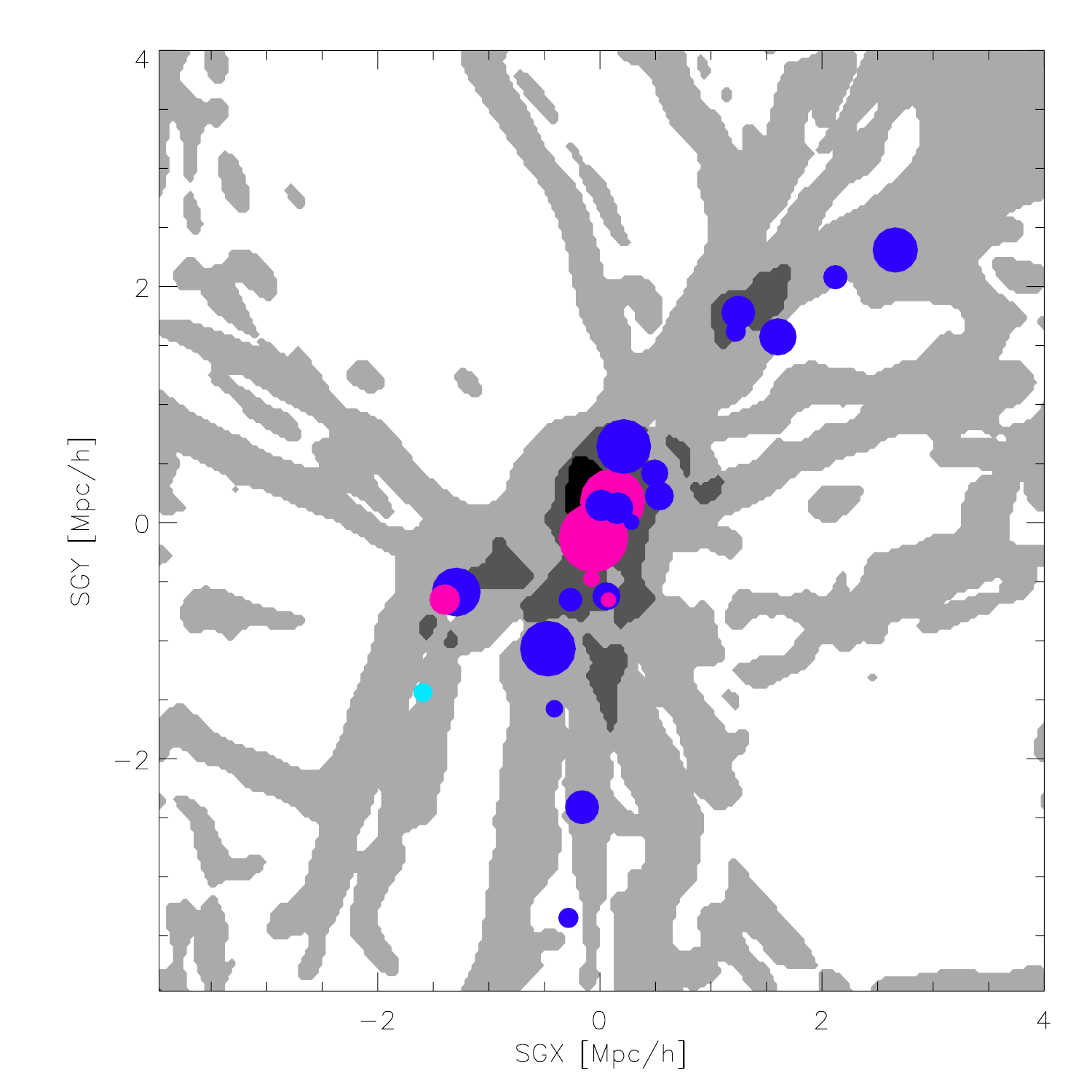}
\includegraphics[width=6.60cm]{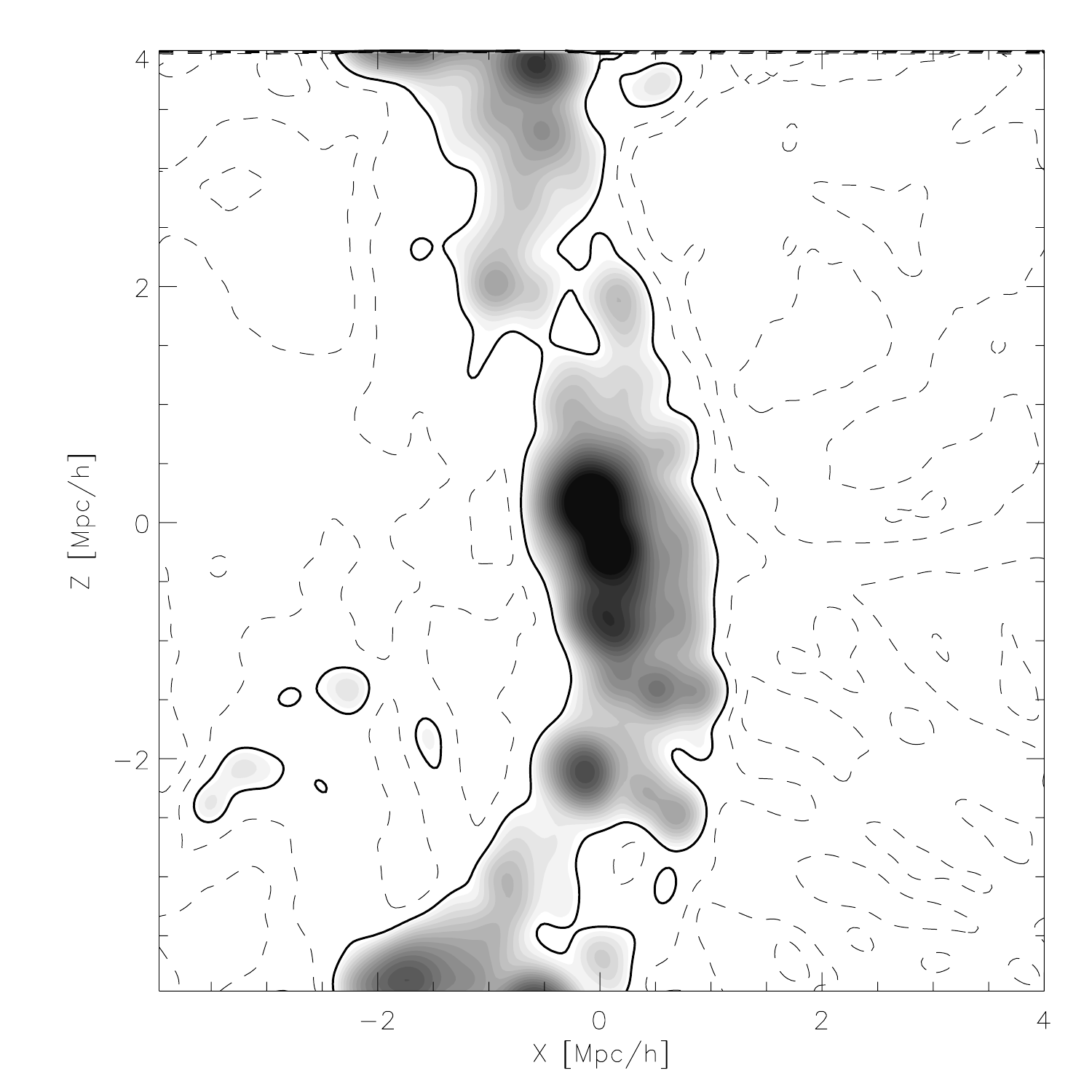}
\includegraphics[width=6.60cm]{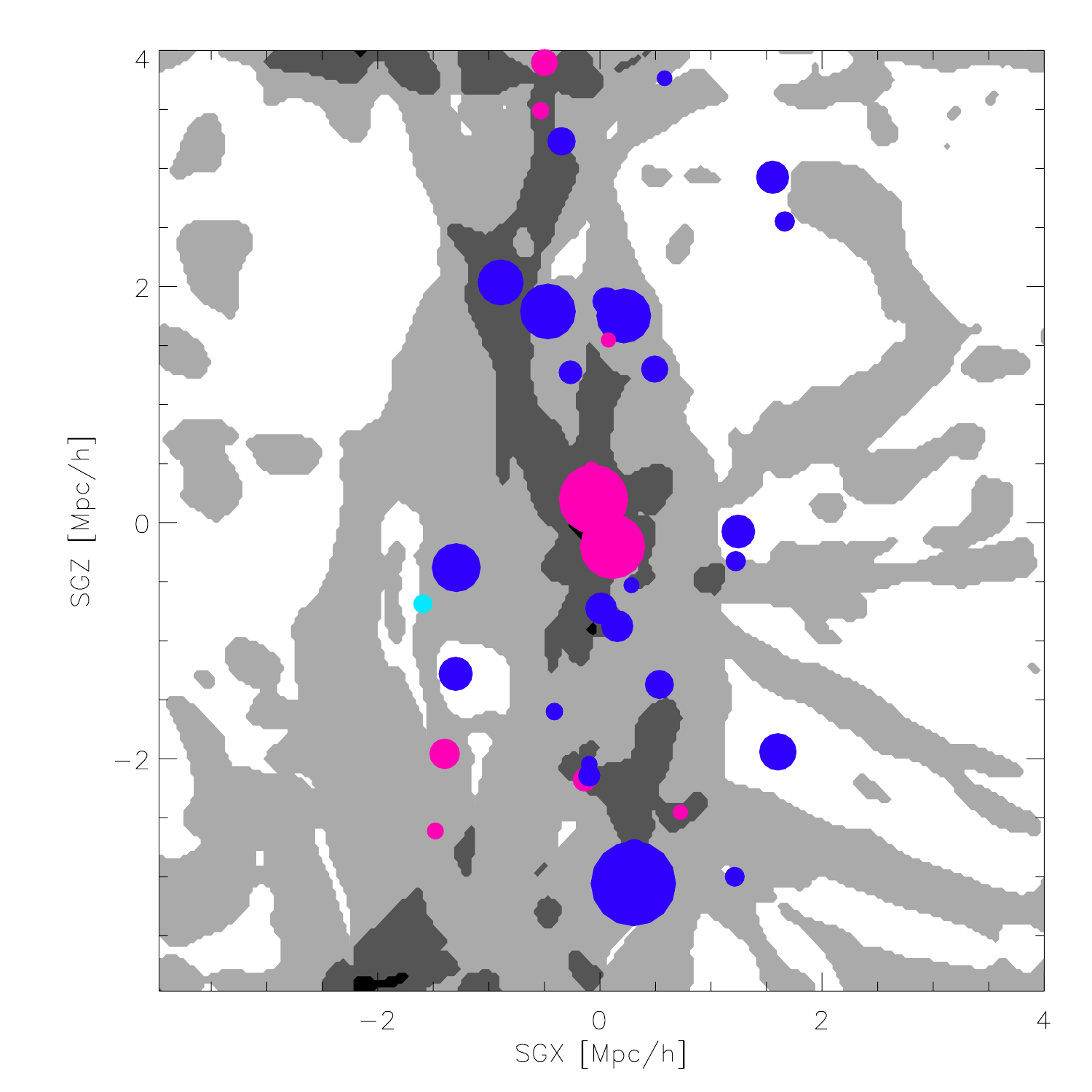}
\includegraphics[width=6.60cm]{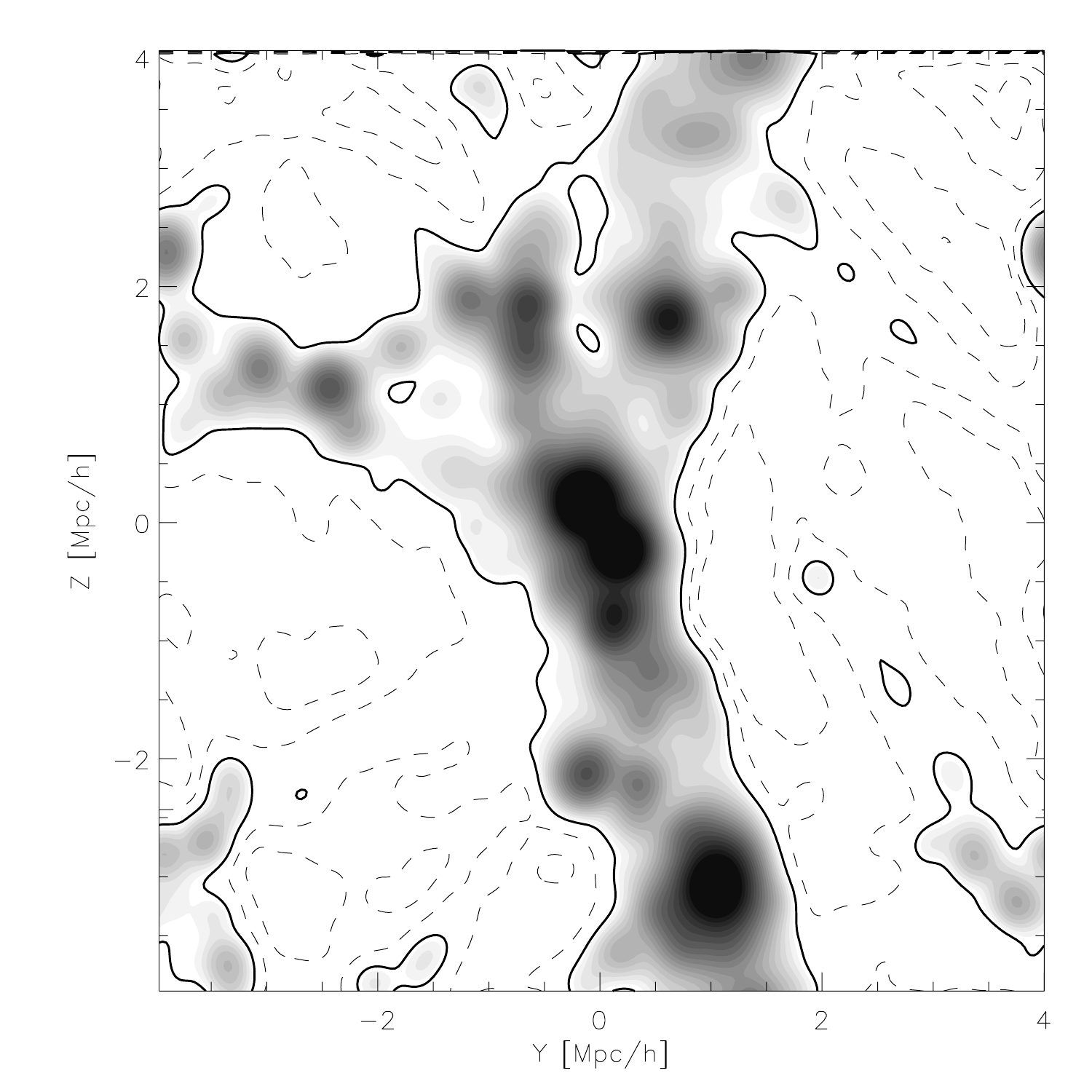}
\includegraphics[width=6.60cm]{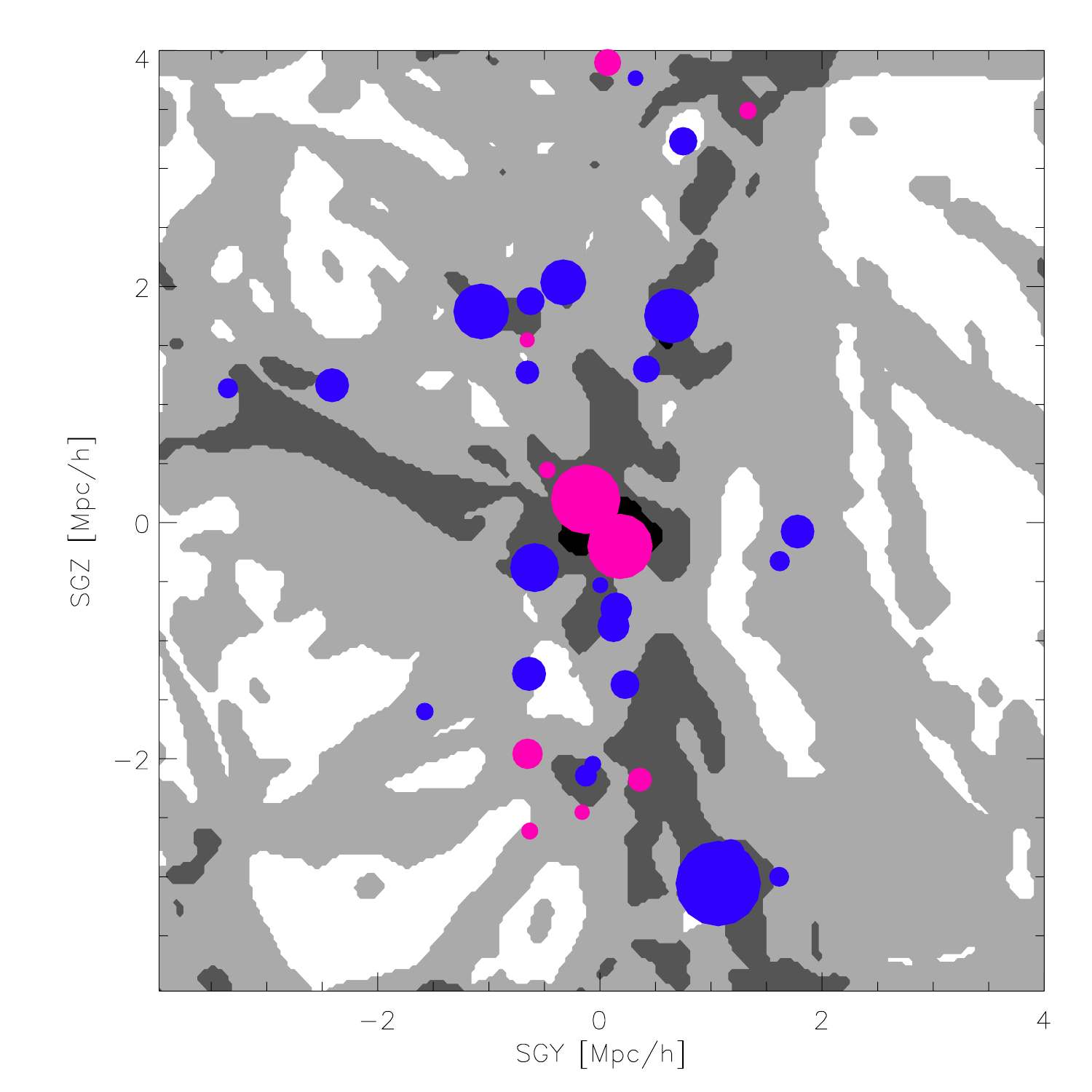}
\caption{ 
A zoom on the simulated LG in the BOX64-5 DM-only simulation. The  logarithm of the fractional density, $\Delta$, (left column) and the    C-web  (right column) of the three   principal planes are shown. 
The   contour coding of Figure \ref {fig:box64-web} is followed here.
The filled coloured circles represent DM halo, with the radii of
circles scaling linearly with the halos' virial radius, and the colours correspond to the web classification  (cyan - sheets, blue - filaments, red - knots).
}
\label{fig:box8_web} 
\end{center} 
\end{figure*}

Fig. \ref{fig:box8_web} zooms in  the inner part of the
computational box and shows the density field, the cosmic web
and the DM halos in the immediate vicinity  of the simulated LG. The plot
shows the three principal Supergalactic planes within a box of a
side length of $8 \hmpc$ centred on the LG. The filled circles
represent the DM halos, with colour representing the web
classification and the size of symbol scaling with halo mass.
The two most massive halos, located at the centre of the zoom
box, are the simulated MW and M31 halos. At the centre of the
box there is a small, i.e. an effective radius of roughly
$\approx 0.5 \hmpc$, blob classified as a knot. This is the
numerical counterpart of the LG, with the centre of the
simulated MW and M31 halos lying just outside the knot, but
having a good part of their mass embedded in the knot. The
simulated group is located within a filament that runs
perpendicular to the Supergalactic Plane, and is close to
coincide with the SGZ axis. As described above,  that filament is
embedded within a sheet that is also running perpendicular
 to the Supergalactic Plane.

\section{The Local Universe as a Dark Matter Laboratory}
\label{lu}

\S \ref{sec:lgwdm}   shows    %
how observations of galaxies in the LG can be used to constrain the nature of DM.   The number of low mass satellites of MW and M31 as compared with CDM predictions  can be explained  through the effects of gasdynamics on baryons, making them invisible, or they can simply not exists, if DM particles  have a mass in the keV scale. 
The nature of the DM affects also the abundance and distribution of isolated dwarf galaxies and the 
 Local Universe provides the optimal test site for confronting  models with observations of such galaxies. The present section presents the comparison of   predictions based on CDM and WDM full box  Box64-3  CLUES simulations (Table \ref{tab:fullbox}) with observations of the distribution of dwarf galaxies.   

\begin{figure}[t]
\begin{center}
\includegraphics[height=6cm]{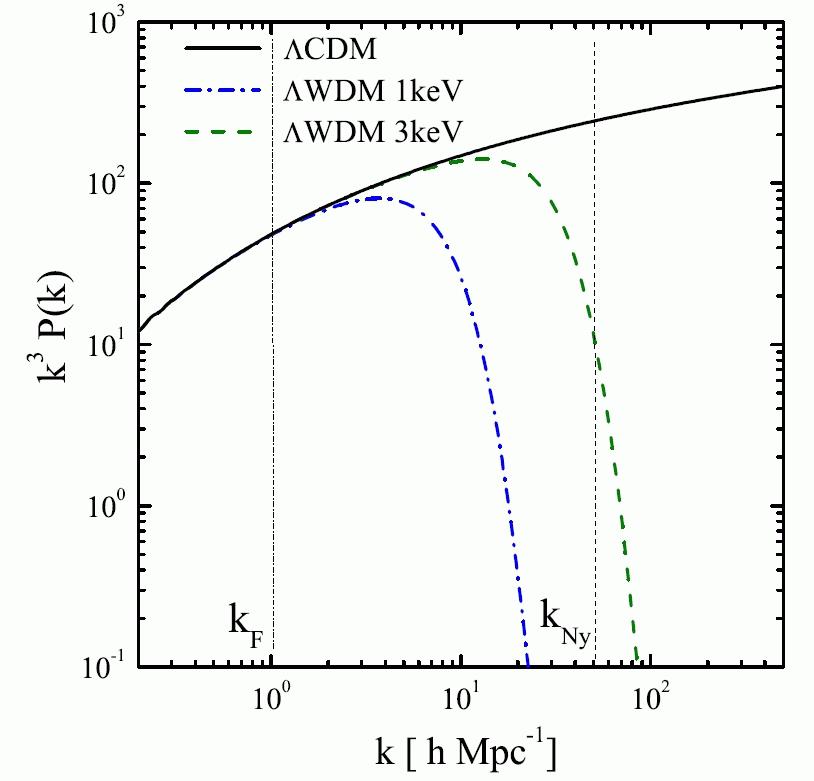}
\caption{ Dimensionless power spectra of a cosmological model with CDM (solid black line)
  and WDM (blue and green  dashed lines). (Figure taken from \citet{2009MNRAS.399.1611T})}
\label{fig:CDM_WDM_PK}
\end{center}
\end{figure}

Fig. \ref{fig:CDM_WDM_PK} shows the underlying primordial power spectrum of the  $\Lambda$CDM and $\Lambda$WDM models. Two cases of the WDM scenario are show, with a mass of the DM particles of 1 and 3 kev. 
We assumed a very low dark matter mass of 1 keV which is a lower
bound set by  observations 
\citep[see the discussion in][]{2009ApJ...700.1779Z, 2009MNRAS.395.1915T}. 
The 1 kev mass corresponds to the maximal impact the WDM model can have on structure formation. 
Same initial conditions are used in the CDM and WDM simulations, modulated by their different power spectra.

\subsection{The size of mini-voids in the Local Volume}

\begin{figure}[ht]
\begin{center}
\includegraphics[width=0.8\textwidth]{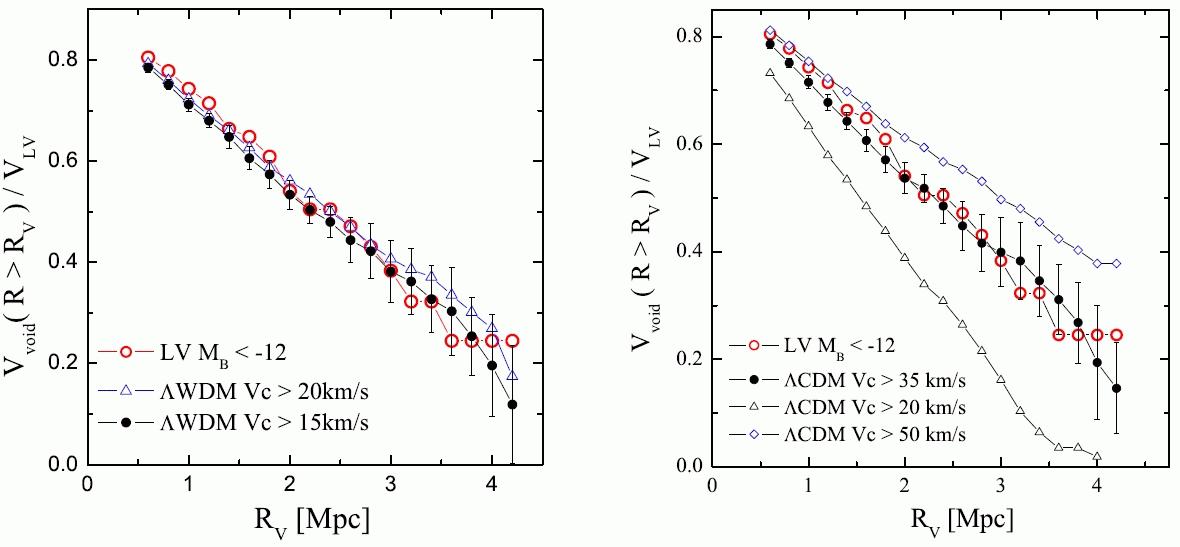}
\caption{ Right: The spectrum of mini-voids in the observational sample with
  $M_B<-12$) (red circles) is compared with the spectrum of mini-voids in a
  halo sample with circular velocity $V_c > 20 \kms$ (open triangles), $V_c >
  50 \kms$ (open diamonds), and $V_c > 35\kms$ (filled black circles) obtained
  from  Box64-3 CDM simulation.  Left: The same in  the Box64-3  WDM simulation but for a halo
  sample with circular velocity $V_c > 20 \kms$ (open blue triangles) and $V_c
  > 15\kms$ (filled black circles). (Figures  taken from \citet{2009MNRAS.399.1611T})}
\label{fig:CDM_WDM_voids}
\end{center}
\end{figure}

The Hubble Space Telescope
observations have provided distances to many nearby galaxies which are 
measured  using the  tip of the Red Giant Branch (TRGB) stars.  Special searches
for new nearby dwarf galaxies have been undertaken by
\citet{2004AJ....127.2031K}. The distances of these galaxies in the Local
Volume are measured independently of redshifts. Therefore, we know both their
true 3D spatial distribution and their radial velocities. The distances have
been measured with accuracies as good as $8-10$\%.  \citet{2009MNRAS.395.1915T}
have used these observational data to construct the spectrum of observed
mini-voids in the Local Volume ($\sim 10$ Mpc around LG). 
They came to the conclusion that the
observed spectrum of mini-voids can be only explained if one assumes that
objects with $V_{c} > 35 \kms $ define the local mini-voids and these objects
host galaxies brighter than $M_B = -12$ (see the right hand side of
Fig. \ref{fig:CDM_WDM_voids}).   
The CDM CLUES simulation predicts almost 500 haloes with
$20 < V_c < 35 \kms$ within the mini-voids in the Local Volume. However only
10 quite isolated dwarf galaxies have been observed with magnitudes $-11.8 >
M_B > -13.3$ and rotational velocities $V_{rot} < 35 \kms$ which points to a
similar discrepancy as the well known predicted overabundance of satellites.
In the WDM model, on the other hand, the truncated power spectrum of the WDM model suppresses the formation of galaxies in small mass halos. 
Thus,  in the WDM CLUES simulation,  the observed spectrum of
mini-voids can be naturally explained if DM haloes with circular
velocities larger than $\sim 15-20 \kms$ host galaxies, as can be clearly seen in  the left panel of  Fig. \ref{fig:CDM_WDM_voids}.  The interested reader  is referred to  \citet{2009MNRAS.399.1611T} for further  details. 

 \begin{figure}[h]
\begin{center}
\includegraphics[height=7cm]{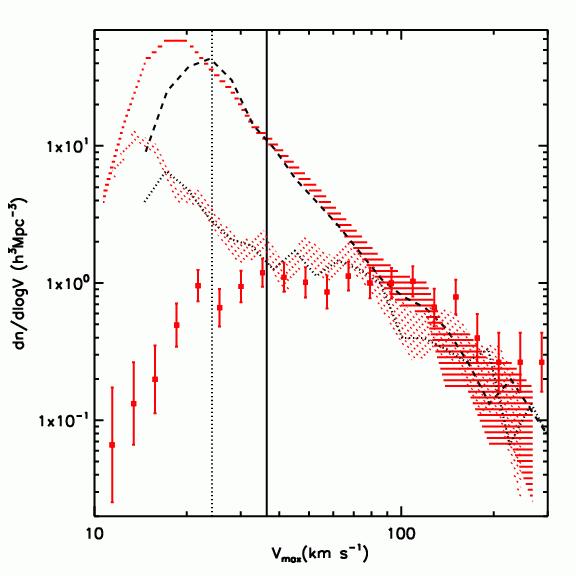}
\caption{The velocity function of ALFALFA galaxies (square symbols with error
  bars).  Predictions from 
 the CLUES simulations for 
  the same field of view are shown  as  the  dashed
  ($\Lambda$CDM)  and dotted  ($\Lambda$WDM) red  areas.
  The vertical solid line marks the value of $V_{max}$ down to which
  the simulations and observations are both complete. (Figure taken from \citet{2009ApJ...700.1779Z})}
\label{fig:ALFALFA}
\end{center}
\end{figure}

\subsection{The abundance of  HI galaxies in the Local Universe}

\citet{2009ApJ...700.1779Z}   compared the velocity function measured from the
Arecibo Legacy Fast ALFA (ALFALFA) survey \citep{2005AJ....130.2613G} with the
velocity function derived from  the Box64-3 CDM and WDM  CLUES simulations.  
Fig. \ref{fig:ALFALFA} shows    that ALFALFA data exhibits a flattening in the velocity function for low circular velocity galaxies that agrees very well with the predictions of Box64-3 WDM CLUES simulation, while the  CDM model present a discrepancy of more than an order of magnitude due to the  over-abundance of low velocity halos.
This results has been recently confirmed  by \citet{2011ApJ...739...38P} using an updated, more complete version of the ALFALFA catalogue.

The fact that much more low mass DM halos are predicted by
cosmological simulations than low luminosity galaxies are observed can  be
explained by gas-dynamical processes which prevent star formation in low mass
halos, as we have explained in \S \ref{sec:lgwdm}.
 Such  gas stripping processes are assumed to occur  in  low mass satellite
halos that are  inside a more massive parent halo. The fact that also dwarfs in the
field are  missing (as described in \citet{2009MNRAS.399.1611T}) is difficult to
explain by gas stripping  induced  by the interaction with their  host halos. 
In  \citet{2013ApJ...763L..41B} we have shown that a dwarf galaxy moving  with high relative speed through a sheet or a filament of the cosmic web is losing a significant fraction of its gas. 
This new mechanism  offers an interesting explanation for  the missing dwarf  galaxy  problem in the Local Volume.

\subsection{Gamma rays from  dark matter annihilation/decay}

One of the main motivations  for launching the  FERMI satellite, that is  scanning the whole sky in the gamma ray band, was   the possible identification of the  nature of DM, assuming it is made of WIMPs.   In that case,  gamma-rays are generated as secondary products of WIMP decay or annihilation  \citep[e.g.][]{strigari}.   In the first case, the  gamma production is proportional to the DM density, while in the case of annihilation, it is proportional to the  density squared.  
This would have made the MW centre to be the prime target for looking for DM related signal. However,  the galactic centre  harbours other sources, of more astrophysical nature, of gamma rays emission, thereby rendering the possible DM-related gamma ray signal very difficult to detect.
Other  targets to look for indirect signal of  DM are the MW satellites and  M31.   However, no  gamma emission from  any of the MW satellites has been detected yet  by FERMI.  
Other nearby DM dominated  structures, like the Virgo and Coma  clusters are the next possible targets for DM detection.
This prompted  \citet{cuesta2009}  to use the CLUES BOX160 DM-only simulation to compute 
 all-sky simulated FERMI maps of gamma-rays from   DM  decay and annihilation in the Local Universe, 
\citet{cuesta2009}  concluded   that FERMI observations of nearby clusters  and filaments are expected to give stronger constraints on decaying DM compared to previous studies.   It was shown, for the first time, that the  filaments of DM distribution in the Local Universe  are  promising targets for indirect detection.  
 On the other hand, the prospects for detection of DM annihilating signal   from nearby structures are less optimistic even with extreme cross-sections.  
 
 Another example of how  the simulations of  the Local Universe can be used  to  set constrains to the physical properties of the possible candidates of DM is shown in \citet{gomez2012}.  
 Using the BOX160 simulation these authors constructed a simulated 
 whole sky  density maps and studied  the prospects of the FERMI telescope to detect  a  monochromatic line of gamma emission due to gravitino decay. The DM halo around the Virgo galaxy cluster was  selected as a reference case, since it is associated with  a particularly high S/N  ratio and is located in a region scarcely affected by the astrophysical diffuse emission from the galactic plane. These authors  found that  a gravitino with a mass range of $0.6-2 $ GeV, and with a lifetime range of about $3 \times 10^{27}-2 \times 10^{28} $ sec  would be detectable by the FERMI with a S/N  ratio larger than 3.  They also  obtained  that gravitino masses larger than about 4 GeV are already excluded  by FERMI data of the galactic halo.

\section{Summary}
\label{summary}

According to the 
 standard model 
of cosmology the universe consists mainly of
Dark Energy and Dark Matter with a small contribution of baryons. The DM  density is about 6 times larger than the baryon density.
DM is the main driver of structure  formation   in the universe, but unfortunately little is known about the nature of the particles which make up the DM. The Local Universe is a very well studied and  observed region in which structures on all scales from clusters of galaxies down to the tiniest   dwarf galaxies can be observed, and thereby it provides the best possibility 
for studying DM  in an astrophysical context.

The CLUES constitutes a framework for performing numerical cosmological simulation that are constrained to reproduced the Local Universe. ``Local" is used here to denote the neighbourhood of the Local Group, extending out to  a depth that ranges typically from a very few to a few tens of Megaparsecs. The CLUES is the numerical counterpart of the Near-Field Cosmology, which aims at studying  cosmology at large by observing the Local Universe and confronting theories and models of structure formation with local findings.  It  constitutes a numerical laboratory designed to experiment with structure formation processes in a way that enables a direct confrontation with the observed Local Universe. 
The current paper is a report on the first steps taken in this direction. 

One of the most exciting challenges that  Near-Field Cosmology and   the CLUES project are facing is   the question of how typical the LG is. To the extent that it can be defined as 'typical' then the study of the LG can shed light on the formation of structure in the universe at large, thus making the study of the near field indeed part of cosmology. In a first attempt to address the problem \citet{2011MNRAS.417.1434F} studied the MAH of simulated LGs, chosen so as  to reproduce the main dynamical features of the LG and its environment, and compared it with the MAH of similar objects selected from the random BOLSHOI simulation. The main finding of the study is that the simulated LGs have a quiet MAH, with their characteristic mass aggregation look-back times  being of the order of 10 Gyrs and thereby statistically significantly longer than those of the random control sample. The interesting point of the MAH study is that the constraints imposed on the initial conditions and on the selection of the LG-like simulated objects are all expressed by present epoch observables. Yet, the LGs that emerge from the simulations are characterized by a long quiescent look-back time, unlike their randomly chosen counterparts. These are preliminary results but they open an interesting window into the mass aggregation history of the LG.

The CLUES simulations have also  been used as a  DM laboratory. This has been conducted along two different paths. One is the study of the formation of a LG-like object  in two cosmological models with CDM and WDM  particles. Using same initial conditions, numerical resolution and sub-grid physics models the CDM and WDM simulations were performed  and compared. \citet{wdmlgrev} have recently presented a detailed analysis of the CDM and WDM constrained simulations. Apart from the expected differences with respect to the abundance and distribution of satellites, and their associated baryonic physics, a new interesting result has been found. 
The  two simulated LG-like objects  are in different stages of evolution. The CDM LG  is beyond its turn-around phase and is more compact. The WDM object, on the other hand   is dynamically  younger, more diffuse and has not reached turned-around. The interesting aspect of the difference between the two cases is that in spite of the fact that the CDM and WDM power spectra coincide on the scale of the LG, namely for mass scale larger  than $\approx 10^{12} \hmpc$, they do differ dynamically. It follows that the cross-talk between different scales affects the dynamics on the LG scale. This does not mean,  of course, that a proper LG-like object cannot be found in a     WDM   scenario.  But, compared with CDM, it would be less likely to find  such an object  in  an environement constrained to mimic the one in which the LG  formed.

The other approach to using the Local Universe as a DM laboratory is by using full box DM-only constrained simulations, assume some simplified model for populating DM halos with galaxies, and then compare the outcome with local surveys of galaxies. By varying the nature of the DM, and hence the power spectrum of the initial condition, while keeping all other aspects of the simulations intact, one can set stringent constraints on the nature of the DM or the model used to associate galaxies with halos. This is the approach adopted by  \citet{2009ApJ...700.1779Z}) and by \citet{2009MNRAS.399.1611T}. Using, what might be considered  a very naive model of associating galaxies with DM halos, the comparison with the observed velocity function of the ALFALFA galaxies \citep{2005AJ....130.2613G} and the spectrum of mini-voids in the very Local Universe \citep{2009MNRAS.395.1915T} clearly favours the WDM model on the CDM one.

CLUES provides a promising way for testing models of galaxy and structure formation within the very Local Universe. The work reported in this paper constitutes only the first steps in that direction. Improved data and more advanced methods of the reconstruction of the primordial initial conditions 
\citep{2013MNRAS.430..912D,2013MNRAS.430..888D,kitaura2013,jenny13}
that seeded the local observed structure are currently being employed by the collaboration. 
 This will result in improved and more tightly constrained simulations which will enable a more thorough experimentation with Near-Field Cosmology in the CLUES numerical laboratory.

\appendix
\section{Overview of CLUES simulations}

The CLUES collaboration has performed a series of numerical simulations of the
evolution of the Local Universe and,  in particular, the Local Group. These
simulations are performed in different boxes and with different
resolutions. There are Dark Matter only simulations as well as simulations with
full gas physics,  including cooling, UV photoionization, star formation, Supernovae feedback and galactic winds. 
A  more general overview can be found at the CLUES web page  ({\tt http://clues-project.org}). These simulations are publicly  available on request. In the following two tables we briefly summarize  the simulations performed in a  full box (Table \ref{tab:fullbox}) as well as
the high-resolution, zoomed-in  resimulations of LG objects (Table \ref{tab:resimu}).

\begin{table}[h!t]
\begin{center}
\begin{tabular}{l|c|c|c|c}
\hline
\hline
Name of simulation  & cosmological  & box size   & number  & particle mass
\tabularnewline
&    model           & $\hMpc$ &  of particles     & $\hMsun$
\tabularnewline
\hline
Box160             &  WMAP3           & 160   &$1024^3$   & $ 2.55 \times10^{8}$  \tabularnewline
Box64-3           & WMAP3           &    64   & $1024^3$   &  $1.64 \times10^{7}$  \tabularnewline
Box64-3-WDM &  WMPA3          &    64    & $1024^3$  &  $ 1.64 \times10^{7}$   \tabularnewline
Box64-5           & WMAP5         &    64   &  $1024^3$    &   $ 1.84 \times10^{7}$   \tabularnewline
\hline
\hline
\end{tabular}
\end{center}
\caption{Full box dark matter only simulations performed within the CLUES
  project.  WMAP3 refers to the
  following set of parameters: $\Omega_{\Lambda} =  0.76$,  $\Omega_{\rm matter}
  =0.24$, $\Omega_{\rm Baryons} =  0.0418$, Hubble parameter $h= 0.73$,
  normalization $\sigma_8 = 0.75$, slope of the primordial power spectrum $n =
  0.95$. WMAP5 refers to the
  following set of parameters: $\Omega_{\Lambda} =  0.721$,  $\Omega_{\rm matter}
  =0.279$, $\Omega_{\rm Baryons} =  0.046$, Hubble parameter $h= 0.7$,
  normalization $\sigma_8 = 0.817$, slope of the primordial power spectrum $n =
  0.96$. 
}\label{tab:fullbox}
\end{table}

\begin{table}[ht]
\begin{center}
\begin{tabular}{l|c|c|c|c}
\hline
\hline
Name of simulation  & cosmological  & type    & number &
DM/stellar particle mass
\tabularnewline
&   model            & $\hMpc$ &    of particles   & $\hMsun$
\tabularnewline
\hline
LG64-3           & WMAP3           &    DM   & $4096^3$   &  $2.55 \times10^{5}$  \tabularnewline
LG64-3           & WMAP3           &    SPH   & $4096^3$   &  $2.22 \times10^{4}$  \tabularnewline
LG64-3-WDM &  WMPA3          &    DM  & $4096^3$  &  $ 2.55 \times10^{5}$   \tabularnewline
LG64-3-WDM &  WMPA3          &    SPH   & $4096^3$  &  $ 2.22 \times10^{4}$   \tabularnewline
LG64-5           & WMAP5         &    DM   &  $4096^3$    &   $ 2.87 \times10^{5}$   \tabularnewline
LG64-5           & WMAP5         &    SPH   &  $4096^3$    &   $ 2.37 \times10^{4}$   \tabularnewline
\hline
\hline
\end{tabular}
\end{center}
\caption{Re-simulations of the the Local Group identified in the simulations
  (in the box of 64 \hMpc\ size) 
  listed in Table \ref{tab:fullbox}. The resimulations are performed either as
  dark matter only simulations (marked as DM) or with hydrodynamics including
  cooling, star formation and feedback (marked as SPH). WMAP3 and WMAP5 refers
  to the set of parameters given in Table \ref{tab:fullbox}. The number of
  particles refers to the formal resolution in the resimulation area (a sphere
  of radius 2 \hMpc\ centred  at the Local Group. The last column provides the
  mass of a DM particle (in DM only simulations) or a stellar  particle (in SPH
  simulations) respectively.
}\label{tab:resimu}
\end{table}


\section*{Acknowledgments}

%

We are very grateful to our CLUES colleagues ({\tt http://clues-project.org}) for their   productive and fruitful collaboration. Without them, this  review  
would have  been  impossible.
GY   thanks the   Spanish\rq{}s  MINECO and MICINN  for supporting his research through different  projects:  AYA2009-13875-C03-02, FPA2009-08958, AYA2012-31101, FPA2012-34694 and Consolider Ingenio SyeC CSD2007-0050. He also  acknowledge support from the Comunidad de Madrid through the ASTROMADRID PRICIT project (S2009/ESP-1496).
YH acknowledges the support of  the  Israel Science Foundation (ISF 1013/13).
SG and YH have been partially supported by the Deutsche Forschungsgemeinschaft under the grant $\rm{GO}563/21-1$.  
We also  thank  the Spanish   MULTIDARK  Consolider project (CSD2009-0006)  and the Schonbrunn Fellowship at the Hebrew  University Jerusalem  for
supporting our collaboration.

The simulations described here have been performed on different supercomputers
at the Leibniz Rechenzentrum Munich (LRZ), the Barcelona Supercomputer Center
(BSC) and the Juelich Supercomputing Center (JSC).

\bibliographystyle{./model2-names}

\bibliography{./refs_all}

\end{document}